%% file: main.tex
\newif\ifdraft
\draftfalse

\newif\ifarxiv
\newif\ifone

\onefalse

\arxivtrue

\ifone
\documentclass[journal,12pt,onecolumn,draftclsnofoot]{IEEEtran}
\else 
\documentclass[lettersize,journal]{IEEEtran}
\fi

\ifarxiv
\else
\fi

\usepackage[utf8]{inputenc}
\input{acro}
\usepackage{units}
\usepackage{url}
\usepackage{cite}
\usepackage{amsmath,amssymb,amsfonts,amsthm}
\usepackage{stackengine}
\stackMath
\usepackage{array}
\usepackage{graphicx}
\usepackage[caption=false,font=normalsize,labelfont=sf,textfont=sf]{subfig}
\usepackage{textcomp}
\usepackage{stfloats}
\usepackage{verbatim}
\usepackage{color}

\newcommand{\stkout}[1]{\ifmmode\text{\sout{\ensuremath{#1}}}\else\sout{#1}\fi}
\ifdraft
\newcommand{\added}[1]{\textcolor{blue}{#1}}
\newcommand{\deleted}[1]{\textcolor{blue}{\stkout{#1}}}

\newcommand{\deletedfloat}[1]{\textcolor{blue}{#1}}
\newcommand{\cmm}[1]{\textcolor{red}{#1}}
\else
\newcommand{\added}[1]{#1}
\newcommand{\deleted}[1]{}

\newcommand{\deletedfloat}[1]{}
\newcommand{\cmm}[1]{}
\fi

\newtheorem*{rep@thrm}{\rep@title}
\newcommand{\newreptheorem}[2]{%
\newenvironment{rep#1}[1]{%
 \def\rep@title{#2 \ref{##1}}%
 \begin{rep@thrm}}%
 {\end{rep@thrm}}}
\makeatother

\newreptheorem{theorem}{Theorem}

\begin{document}

\title{Distributed Machine-Learning for Early HARQ Feedback Prediction in Cloud RANs}
\author{Bar{\i}{\c s} G{\" o}ktepe,\ifarxiv~\fi
Cornelius Hellge,\ifarxiv~\fi
Thomas Schierl,\ifarxiv~\fi
and Slawomir Stanczak
\thanks{All authors are with Fraunhofer Heinrich Hertz Institute, 10587 Berlin, Germany (e-mail: firstname.lastname@hhi.fraunhofer.de). Slawomir Stanczak is also associated with the Technical University of Berlin.
\ifarxiv
\newline © 2023 IEEE. Personal use of this material is permitted. Permission from IEEE must be obtained for all other uses, in any current or future media, including reprinting/republishing this material for advertising or promotional purposes, creating new collective works, for resale or redistribution to servers or lists, or reuse of any copyrighted component of this work in other works.\fi}%
\thanks{Manuscript received October 21, 2022; revised January 07, 2023; accepted April 27, 2023.}}

\ifarxiv
\IEEEpubid{10.1109/TWC.2023.3275296~\copyright~2023 IEEE}
\fi

\maketitle

\begin{abstract}
     In this work, we propose novel HARQ prediction schemes for Cloud RANs (C-RANs) that use feedback over a rate-limited feedback channel (2 - 6 bits) from the Remote Radio Heads (RRHs) to predict at the User Equipment (UE) the decoding outcome at the BaseBand Unit (BBU) ahead of actual decoding. In particular, we propose a Dual Autoencoding 2-Stage Gaussian Mixture Model (DA2SGMM) that is trained in an end-to-end fashion over the whole C-RAN setup. Using realistic link-level simulations in the sub-THz band at 100 GHz, we show that the novel DA2SGMM HARQ prediction scheme clearly outperforms all other adapted and state-of-the-art schemes. The DA2SGMM shows a superior performance in terms of blockage detection as well as HARQ prediction in the no-blockage and single-blockage cases. In particular, the DA2SGMM with 4~bit feedback achieves a more than 200~\% higher throughput in average compared to its best alternative. Compared to regular HARQ, the DA2SGMM reduces the maximum transmission latency by more than 72.4~\%, while maintaining more than 75~\% of the throughput in the no-blockage scenario. In the single-blockage scenario, DA2SGMM significantly increases the throughput for most of the evaluated Signal-to-Noise-Ratios (SNRs) compared to regular HARQ.
\end{abstract}

\begin{IEEEkeywords}
Early HARQ, feedback, prediction, machine learning, Cloud RAN.
\end{IEEEkeywords}

\section{Introduction}
\IEEEPARstart{T}{he} emergence of new services, such as \ac{V2X}, \ac{VR}, \ac{URLLC} and many more, has increased the need\deleted{s} for higher data rates and extremely low latencies. This has directed the interest of mobile communication standards to cover new and higher frequency bands. The \ac{5G} standardization body, the \ac{3GPP}, has recently finished a new work item targeting frequencies up to \unit[71]{GHz} for access \cite{71_ghz_wid}. In particular, recent advances in hardware have paved the way for using these bands. The sub-THz and THz bands, which reach from \unit[100]{GHz} up to \unit[3]{THz}, are now in the focus for beyond \ac{5G} technologies \cite{8732419, dbbs_mods_00068882}. However, the use of high\added{-}frequency bands has the disadvantage of being highly dependent on an unobstructed \ac{LOS} path\deleted{,} and having significantly shorter channel coherence times, which require a higher control signaling overhead due to more frequent channel measurements \cite{8732419}. Especially, the latter poses a bottleneck for the \ac{CSIT}, which arrives with a delay \cite{9269936}. \ac{CSIT} is essential to estimate the appropriate transmission parameters, such as the \ac{MCS}, precoding, etc. Especially, for highly mobile \acp{UE}, such as cars or trains, the \ac{CSIT} is already outdated when it is available at the transmitter. Although, exploiting geometrical properties of the environment and employing \ac{ML} enables predicting the \ac{CSIT} over larger time windows \cite{9269936}, the fast fading behavior of the channel may still make the channel estimation inaccurate. 

To cope with inaccurate \ac{CSIT}, physical layer retransmission mechanisms, such as \ac{HARQ}, are used. However, \ac{HARQ}, also known as reactive \ac{HARQ}, increases the end-to-end latency because the transmitter requires feedback from the receiver after each transmission round in form of an \ac{ACK} or \ac{NACK}. Especially, for \ac{5G} \ac{URLLC} use cases with end-to-end latency requirements of down to \unit[1]{ms} \cite{5g_paper}, reactive \ac{HARQ} poses a limitation. For \ac{6G} use cases, where end-to-end latency requirements even down to \unit[100]{\textmu s} are foreseen \cite{9428726}, this becomes even more an issue. This drawback is compensated by proactive \ac{HARQ} that continuously transmits further retransmissions until an \ac{ACK} is received \cite{gf_harq_oulu, gf_harq_aalborg}. Proactive \ac{HARQ} combines high reliability with extremely short latencies \cite{liu2020analyzing}. Nevertheless, it trades these advantages for a degraded spectral efficiency due to unnecessary retransmissions \cite{feedback_prediction_iiot}. \IEEEpubidadjcol 

The dependence on the \ac{LOS} path also poses a major issue for reliable communication, as any obstruction by an object causes a severe degradation of the channel quality. As a remedy, \ac{C-RAN} architectures with multiple reception points at different locations, i.e. \acp{RRH}, are foreseen for sub-THz and THz communications \cite{8782882}. The \ac{BBU}, which is responsible for higher layer processing, decodes the packet by combining all received signals from the different \acp{RRH}. However, in the context of \ac{C-RAN} architectures, the aforementioned drawbacks of reactive \ac{HARQ} and proactive \ac{HARQ} become even more critical due to the significantly larger feedback delay \cite{prediction_cran2}. In particular, a fronthaul latency of up to 250 \textmu s is assumed \cite{ngmn_cran}. Hence, many papers in the scientific literature studied ways for reducing the feedback delay using prediction mechanisms \cite{prediction_cran1, prediction_cran2, caire_fast_harq}. For architectures with a single reception point, different \ac{HARQ} feedback prediction methods exist \cite{caire_fast_harq, zhou_ieeetran, phdthesis_csi_harq, snr_prediction, prediction_cran1, prediction_cran2, deep_ml_eharq_iq, early_harq_schemes2,early_harq_schemes, prediction_nn, llr_channel_harq_prediction, ldpc_subcodes, journal_eharq_paper, feedback_prediction_iiot}. In contrast, for architectures with multiple reception points, i.e. \acp{C-RAN}, we know only \ac{SNR}-based \ac{HARQ} schemes proposed by Khalili and Simeone in \cite{prediction_cran1} and Makki et. al. in \cite{caire_fast_harq}. Other prediction mechanisms, such as \ac{LLR}-based and subcode-based approaches, \cite{feedback_prediction_iiot} and \cite{early_harq_schemes2, early_harq_schemes, prediction_nn, llr_channel_harq_prediction, ldpc_subcodes, journal_eharq_paper}, have not been adapted yet to \ac{C-RAN} architectures. Current state-of-the-art designs assume that the predictor has full knowledge of the prediction features. However, in \ac{C-RAN} architectures, the \acp{RRH} only have partial knowledge and further, the feedback channels to the \ac{UE} are rate-limited. Hence, in \ac{C-RAN} architectures, \ac{HARQ} prediction schemes that consider the locality of the information are required. Due to the rate-limitation of the feedback channels, schemes also have to develop efficient representations of the local feedback and rules on how to combine these. Furthermore, even for the \ac{SNR}-based approach, no evaluation using realistic link-level simulations in the \ac{C-RAN} context, in particular considering blockage, is available. Against this background, the contributions of this paper are summarized in the following:
\begin{itemize}
    \item To address the \ac{HARQ} prediction problem in \acp{C-RAN} in a holistic manner, we present a novel \ac{DIDA} that exploits subcode features as well as channel estimation features. Furthermore, to reduce the dimensionality of the input features we propose over \cite{journal_eharq_paper} and \cite{feedback_prediction_iiot} a subcarrier-based averaging of the \acp{LLR} for the \ac{DIDA}.
    \item To enable the application of state-of-the-art feedback prediction mechanisms for single reception points in \acp{C-RAN}, we propose a distributed \ac{HARQ} prediction setup with quantization of the feedback and a combining rule at the \ac{UE}. Within this setup, we develop a distributed \ac{LR-LLR} based on \cite{llr_channel_harq_prediction}.
    \item Finally, we compare all schemes in the context of our \ac{HARQ} system evaluation methodology using realistic link-level simulations. In particular, we also consider the single-blockage case, where one \ac{RRH} is blocked. We show that the \ac{DIDA} clearly outperforms all other schemes in all experiments.
\end{itemize}

\subsection{Related work on HARQ feedback prediction}
As mentioned in the previous section, different \ac{HARQ} prediction schemes have been proposed and studied in the literature. The variety of schemes reaches from simple thresholding, e.g. \cite{prediction_cran2}, up to complex machine learning schemes, e.g. \cite{prediction_nn} and \cite{journal_eharq_paper}. In particular, the latter has gained interest recently, also in the context of the new Rel.~18 standardization \cite{ai_ml_qc}. In the following, we classify these schemes into three categories:

\begin{enumerate}
    \item \textbf{Channel-estimation-based feedback prediction:} In \cite{caire_fast_harq}, Makki et al. investigated a mixture of proactive and reactive \ac{HARQ} protocols to reduce the expected latency. In the proposed scheme, the receiver accumulates the received signal until the sum channel gain that is estimated over quantization regions, exceeds a certain threshold associated with a sufficiently high decoding probability. In case of a negative prediction, i.e. the transmitted redundancy is not sufficient for successful decoding, the receiver switches to a reactive \ac{HARQ} approach. In \cite{prediction_cran2}, Rost and Prasad and, in \cite{prediction_cran1}, Khalili and Simeone put the channel-estimation-based prediction schemes into the context of \acp{C-RAN} and showed the benefits of early feedback in \acp{C-RAN} with non-ideal backhaul.
    
    Besides that, in \cite{deep_ml_eharq_iq}, AlMarshed et al. proposed a Deep \ac{ML} scheme that uses the received complex signal to estimate the decodability of a packet. Although being different from the previously described schemes, we categorize the Deep ML as a channel-estimation-based approach because it does not involve the computation of \acp{LLR} or any other channel-code-aware features.
    
    \item \textbf{\ac{LLR}-based feedback prediction:} In \cite{early_harq_schemes2} and \cite{early_harq_schemes}, Berardinelli et al. used a \ac{BER} estimate based on \acp{LLR} to predict the decoding outcome ahead of the actual decoding. They empirically computed a threshold for the \ac{BER} estimate to predict the decodability. In contrast to the channel-estimation-based schemes, the \acp{LLR} inherently contain a reduced form of the channel estimates. Nevertheless, different from the first category, the \ac{LLR}-based schemes consider the whole received data signal for the prediction instead of relying only on pilots used for the channel estimation. As an improvement over the simple thresholding that was used by Berardinelli et al., Hummert et al. proposed in \cite{prediction_nn} to use a neural network, designated as NN ForeCast, that can mimic the decoder. A hybrid pathway was proposed by AlMarshed et al. in \cite{llr_channel_harq_prediction}. They combined both \ac{LLR}- and channel-estimation-based features using a logistic regression. The proposed logistic regression showed a significant enhancement over other approaches that use only one of both.
    
    \item \textbf{Subcode-based feedback prediction:} In \cite{feedback_prediction_iiot, ldpc_subcodes} and \cite{journal_eharq_paper}, the authors proposed a feedback prediction mechanism that observes the partial decoding behavior of so-called subcodes. The subcodes reflect dependencies between received symbols arising from the structure of the channel code. Similar to the \ac{LLR}-based feedback prediction, this approach uses the \acp{LLR} as a basis. However, in contrast to these, subcode-based schemes apply additional processing on the \acp{LLR} based on the knowledge of the code structure. In \cite{ldpc_subcodes}, Göktepe et al. empirically determined thresholds for these code-aware features. As an improvement over this thresholding approach, in \cite{journal_eharq_paper}, the authors applied machine learning techniques, i.e. logistic regression, random forests, isolation forests and supervised autoencoders, on the \ac{LLR} and subcode features. Especially, the logistic regression and the supervised autoencoder have proven to be fruitful approaches to enhance the feedback prediction.
\end{enumerate}

Apart from the previously described \ac{HARQ} prediction schemes that are mainly designed for single reception points, \cite{prediction_cran2}, \cite{prediction_cran1} and \cite{caire_fast_harq} presented similar approaches for implementing the channel-estimation-based feedback prediction in \acp{C-RAN}. The works proposed collecting the channel gains and the \acp{SNR} over the received \acp{RV}, respectively. In \cite{prediction_cran2}, Rost and Prasad applied Gallager's error exponent $E_r$ to convert the channel estimation into an estimated error probability $\epsilon$, as this metric is easier to work with (see \cite{gallager_book} for more details):
\begin{eqnarray}
    \epsilon(R, \gamma) &=& c_1 e^{-NE_r(R,\gamma)}\,,\label{eq:gallager}\\
    E_r(R,\gamma) &=& R_0(\gamma) - R\,,
\end{eqnarray}
where $\gamma$ is the average \ac{SNR}, $R$ is the code rate, $N$ is the code length, and $R_0$ is the cut-off rate associated with the \ac{SNR}. After calculating Gallager's error exponent locally at the \acp{RRH}, they proposed applying a threshold to generate a positive or negative feedback. However, they ignored the case of multiple \acp{RRH} receiving the same packet. This was investigated by Khalili and Simeone in \cite{prediction_cran1}. They proposed using a vector quantization, as described in \cite{mimo_vector_quantization}, to compress the channel state at the \acp{RRH}. Following the compression, the \acp{RRH} transmit this compressed feedback over a rate-limited feedback channel to the \ac{UE} where a joint feedback is calculated. Furthermore, they also analyzed the impact of quantization of feedback on the system performance.\\

\emph{Notation:} Throughout the paper, we use $\mathbb{C}$ to denote the set of complex numbers and $\mathbb{N}$ the set of natural numbers. Furthermore, $\mathbb{C}^n, n\in \mathbb{N},$ denotes the $n$-dimensional complex vector space. Bold letters are used to indicate vectors, while bold capital letters are used to indicate matrices. Random variables are noted in capital letters, where random matrices are further highlighted by bold font. $\mathbb{E}[\cdot]$ is the expected value. $\mathrm{diag}(\cdot) : \mathbb{C}^n \to \mathbb{C}^{n \times n}, n \in \mathbb{N},$ denotes the mapping of an $n$-dimensional complex vector to an $n \times n$ complex diagonal matrix where the diagonal elements of the matrix are the components of the vector. $\mathcal{N}(\mu, \sigma^2)$ designates the normal distribution with mean $\mu$ and variance $\sigma^2$ and $\mathcal{CN}(\mu, \sigma^2)$ designates its circularly-symmetric complex counterpart.

\section{System Model}
\begin{figure}[t]
  \centering
  \ifone
  \includegraphics[width=.55\columnwidth]{cloudran_harq}
  \else
  \ifarxiv
  \includegraphics[width=.9\columnwidth]{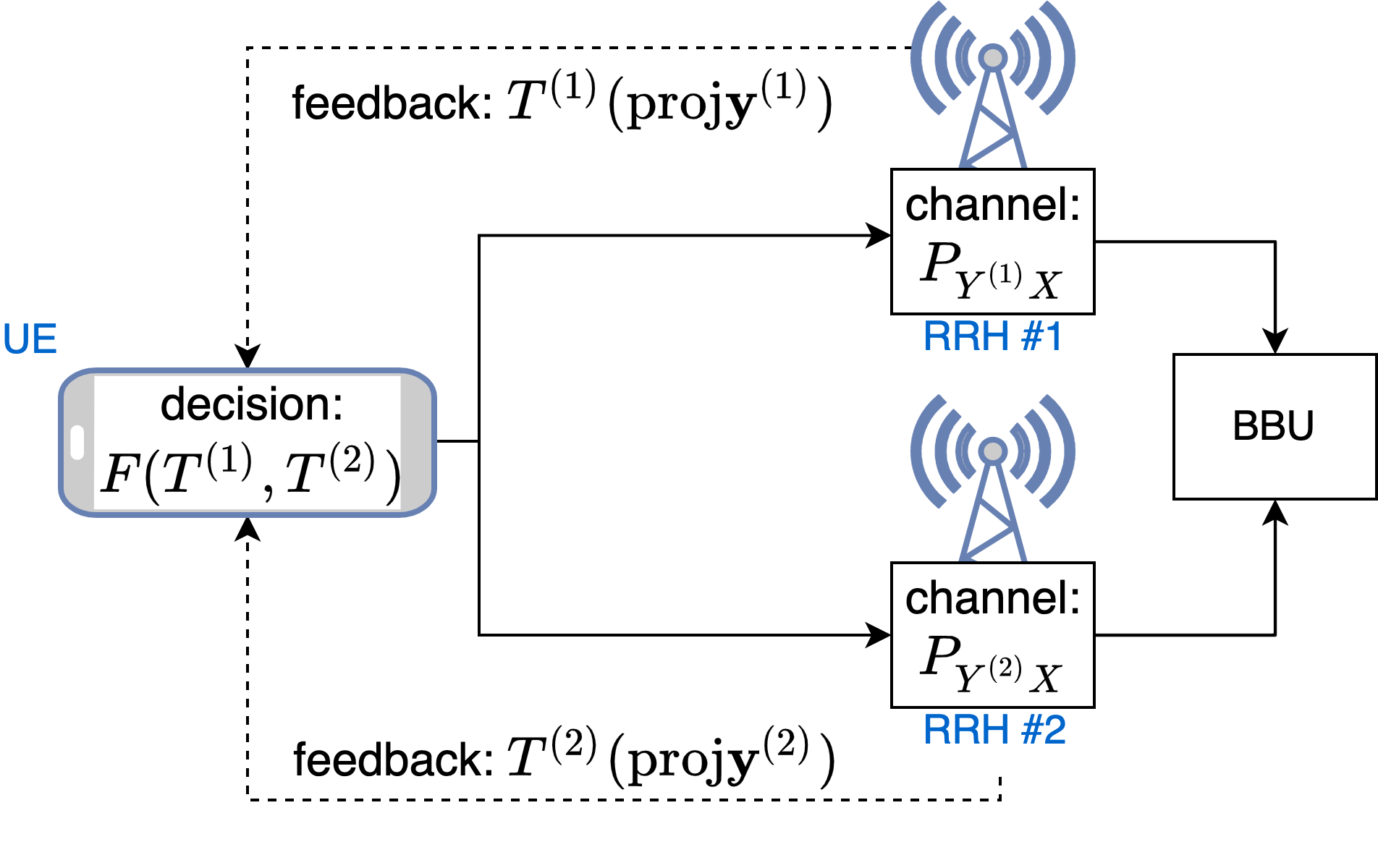}
  \else
  \includegraphics[width=.9\columnwidth]{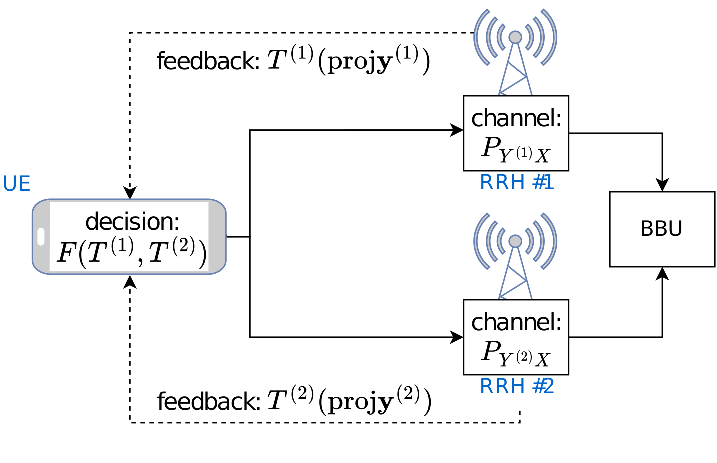}
  \fi
  \fi
  \caption{Uplink C-RAN scenario with local HARQ feedback generated at the RRHs which is combined at the UE.}
  \label{fig:cloudran_setup}
\end{figure}

\begin{table}[t]
\centering
\caption{Definition of parameters}
\label{tab:system_params}
\begin{tabular}{|l|l|}
\hline
Parameter&Definition\\
\hline
$n$ &Number of modulation symbols for the decoding attempt\\
&at the \ac{BBU}\\
$p$ &Number of modulation symbols used at the time of\\
&prediction\\
$N_\mathrm{b}$&Number of bits used for feedback from \ac{RRH} to \ac{UE}\\
$T_\mathrm{max}$&Maximum number of transmissions/\acp{RV}\\
$M$&Number of bits per modulation symbol, i.e. QAM\\
&modulation order\\
$N_\mathrm{sc}$&Number of subcode features\\
\hline
\end{tabular}
\end{table}

Fig.~\ref{fig:cloudran_setup} shows the system setup which is used throughout the paper. The definitions of commonly used variables are summarized in Tab.~\ref{tab:system_params}. We assume an uplink scenario where a \ac{UE} is transmitting a packet, which is simultaneously received by two \acp{RRH}.
After partially receiving a packet, the \acp{RRH} generate a local feedback based on the evaluated prediction algorithms. The generated local feedback is transmitted and combined at the \ac{UE}. In the meanwhile, the received signals from the \acp{RRH} are accumulated and jointly decoded at the \ac{BBU} which determines the final decoding outcome.\\
Let $\mathrm{A} := \mathbb{C}^{n}$ and $\mathrm{B} := \mathrm{B}_1 \times \mathrm{B}_2 = \mathbb{C}^{p} \times \mathbb{C}^{(n-p)} = \mathbb{C}^{n}$ designate the input and output sets. Furthermore, let each channel be characterized by its respective conditional probability measure $P_{Y^{(1)}|X} : \mathrm{A} \mapsto \mathrm{B}$ and $P_{Y^{(2)}|X} : \mathrm{A} \mapsto \mathrm{B}$. Given $n$ modulation symbols, the channel probability measures represent the following association between the random vector representing the transmitted signal $X \in \mathrm{A}$ and the received signal random vectors ${Y}^{(1)}, {Y}^{(2)} \in \mathrm{B}$:
\begin{equation}
{Y}^{(i)} = \boldsymbol{H}^{(i)} X + {Z}^{(i)},\, i = 1,2\,, \label{eq:system_model}
\end{equation}
where $\boldsymbol{H}^{(i)} \in \mathrm{B}$, $i=1,2$, are random fading matrices and ${Z}^{(i)} \in \mathrm{B}$, $i=1,2$, are random noise vectors with each element distributed according to $\mathcal{CN}(0, 1)$. In our link-level simulations, we assume a spatially filtered \ac{CDL} channel model \cite{3gpp_channel_models_100}. The assumed channel model, which is explained more in detail in Sec.~\ref{sec:lls}, can be modeled as complex channel gains that distort each symbol individually and additive normally distributed noise on top.\\
To determine the final decoding outcome, the \ac{BBU} combines and jointly decodes the received signal vectors $Y^{(1)}$ and $Y^{(2)}$ from both \acp{RRH}. On the other hand, the \acp{RRH} calculate the feedback, which is a map $T^{(i)} : \mathrm{B}_1 \to \mathrm{S}$, $i=1,2$, where $\mathrm{S}$ is the sample space of the feedback. $T^{(i)}$ is specified differently depending on the prediction scheme. As we assume binary communication, the feedback sample space reduces to $\mathrm{S} := \mathbb{F}_2^{N_\mathrm{b}}$ where $N_\mathrm{b}$ is the number of bits used for the feedback transmission. Finally, the \ac{UE} applies a combination rule $F : \mathrm{S} \times \mathrm{S} \to \{\mathrm{ACK}, \mathrm{NACK}\}$, which leads to the corresponding \ac{UE} behavior, i.e. stop transmitting or transmit more redundancy.

\section{Distributed early HARQ strategies}
In this paper, we consider early \ac{HARQ} strategies that attempt to predict the decodability of a packet ahead of the actual decoding. In particular, we take only a part of the whole transmitted signal vector into account. In contrast to early \ac{HARQ} strategies that use the whole signal vector, this approach allows for providing the feedback at an earlier stage, which is crucial especially for latency\added{-}constrained use cases. In the broadest sense, the decodability prediction can be interpreted as binary statistical hypothesis testing, where the early \ac{HARQ} predictor tries to discriminate between two probability distributions $P$ and $Q$: the probability distribution of decodables and the probability distribution of undecodables. The feedback maps $T^{(1)}$ and $T^{(2)}$ have to be chosen such that the two distributions become as distinguishable as possible. Ideally, these maps are sufficient statistics to the statistical hypothesis testing problem. However, in practice, depending on the type of feedback, it is a notoriously difficult problem to exactly characterize these distributions and hence, also finding sufficient feedback maps.\\
In terms of the system model, the prediction is based on $p$ modulation symbols with $p < n$. 
Hence, given $\mathrm{proj}_p{Y}^{(1)}$ and $\mathrm{proj}_p{Y}^{(2)}$, the binary hypothesis testing task at the \ac{UE} is to decide between the two distributions 
\begin{eqnarray}
P &:=& P_{(T^{(1)}(\mathrm{proj}_pY^{(1)}),T^{(2)}(\mathrm{proj}_pY^{(2)}))|D=\textrm{ACK}}\\
Q &:=& P_{(T^{(1)}(\mathrm{proj}_pY^{(1)}),T^{(2)}(\mathrm{proj}_pY^{(2)}))|D=\textrm{NACK}}\,,
\end{eqnarray}
where $D \in \{\mathrm{ACK}, \mathrm{NACK}\}$ respresents the decoding outcome at the \ac{BBU} with $n$ modulation symbols available at the decoder and $\mathrm{proj}_p : \mathbb{C}^{k \times p} \times \mathbb{C}^{k \times (n-p)} \mapsto \mathbb{C}^{k \times p}$, $p,k,n \in \mathbb{N}$, $p < n$, denotes the function, which maps an element from the Cartesian product of two vector spaces on the first vector space.

\subsection{Channel-estimation-based HARQ prediction (Q-SNR)}\label{sec:channel_pred}
Channel estimation predictors focus on the estimated channel realization $\boldsymbol{\hat{H}}^{(i)}$, $i=1,2$, at each \ac{RRH}. The estimation is performed based on known parts of the transmitted signal, e.g. reference signals, such as \ac{DMRS}. In the particular case of the paper's transmission model, one \ac{DMRS} is located at the beginning of each \ac{RV}. We use these \ac{DMRS} to obtain the received \acp{SNR}, $\gamma^{(1)}$ and $\gamma^{(2)}$, at each \ac{RRH}, respectively.

\begin{figure}[t]
  \centering
  \ifone
  \includegraphics[width=.55\columnwidth]{distributed_snr}
  \else
  \ifarxiv
  \includegraphics[width=.9\columnwidth]{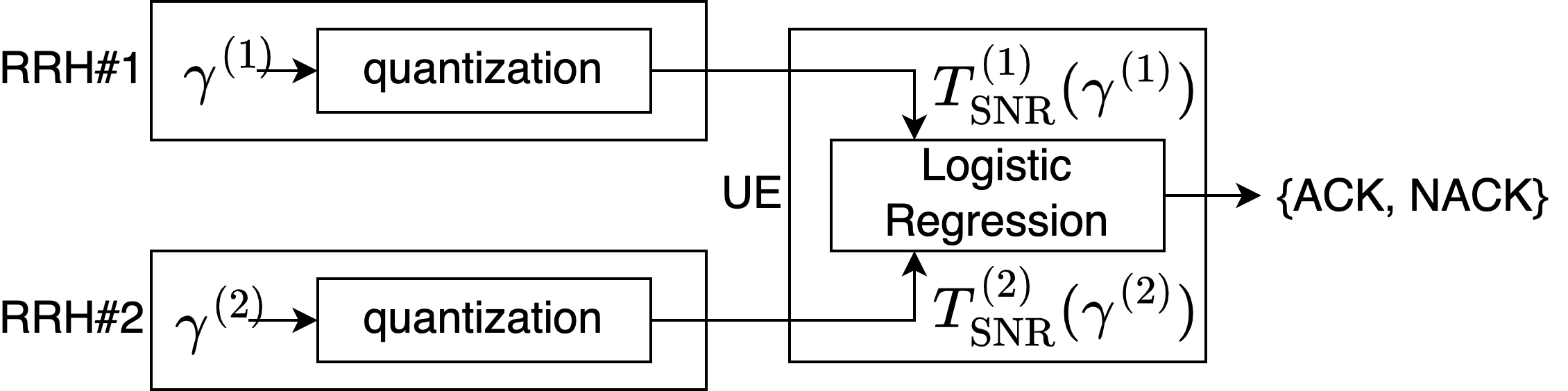}
  \else
  \includegraphics[width=.9\columnwidth]{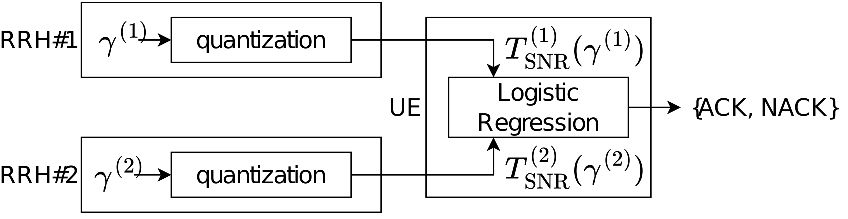}
  \fi
  \fi
  \caption{\ac{SNR} based distributed prediction approach.}
  \label{fig:distributed_snr}
\end{figure}
For the channel-estimation-based prediction, we evaluate the scheme proposed in \cite{caire_fast_harq} that accumulates the quantized received \acp{SNR} from the \acp{RRH} at the \ac{UE} and applies a threshold to the sum to predict an \ac{ACK} or \ac{NACK}. In particular, $F_\mathrm{Q-SNR} := T_\mathrm{SNR}^{(1)} + T_\mathrm{SNR}^{(2)} \stackunder{\stackon{\lessgtr}{\scriptstyle \mathrm{NACK}}}{\scriptstyle \mathrm{ACK}} C_\mathrm{Q-SNR}$, where $T_\mathrm{SNR}^{(i)}$, $i=1,2$, are the quantized received \acp{SNR} from the respective \acp{RRH} and $C_\mathrm{Q-SNR}$ is a constant that controls the trade-off between false-positive and false-negative errors. In the constant power case, the received \ac{SNR} is equivalent to the accumulated channel gain, which is used in \cite{caire_fast_harq}. Furthermore, we model the quantization by a quantization layer. We assume both quantization functions $T_\mathrm{SNR}(\gamma) := T^{(1)}_\mathrm{SNR}(\gamma) = T^{(2)}_\mathrm{SNR}(\gamma) \forall \gamma$ to be the equal and further to be piece-wise constant functions. The constant intervals are chosen, such that each interval contains approximately the same number of data points, over the relevant \ac{SNR} range, where the relevant range is determined by the minimum and maximum values of the training set. See "quantile" in \cite{scikit_kbin} for more details. Then, $T_\mathrm{SNR}$ assigns the value of the interval center to each \ac{SNR} that falls into that interval. This scheme is designated as \ac{Q-SNR} in the following. In \cite{prediction_cran1} and \cite{caire_fast_harq}, the authors use the error probability approximation from \cite[Eq. (59)]{6802432} to estimate the failure probability. In particular\added{,} in \cite{prediction_cran1}, Khalili and Simeone apply a threshold to the estimated error probability. However, in our evaluated scenario, the \ac{Q-SNR} scheme achieves the same performance as the scheme in \cite{prediction_cran1} at a significantly lower complexity. Hence, we restrict only to the \ac{Q-SNR} scheme.

\subsection{LLR-based HARQ prediction (LR-LLR)}
The \ac{LLR}-based approaches assume that each element $x_i \in \mathrm{S} \subset \mathbb{C}$, $i=1,2,...,n$, of the transmitted signal vector $\boldsymbol{x}$ is i.i.d. and each element of the symbol set $\mathrm{S}$ representing $M$ bits has the same probability. We are aware that this assumption does not hold in practice due to the channel code. Nevertheless, in the next section, we discuss a scheme that does not resort to this assumption. Using the i.i.d. assumption, the \acp{LLR} are calculated as 
\begin{equation}\label{eq:llr}
\Lambda_{(q-1)M + j}^{{(i)}} := \log\frac{P(b_{(q-1)M + j} = 1 | r_{q})}{P(b_{(q-1)M + j} = 0 | r_{q})}\,,\,i=1,2\,,
\end{equation}
where $r_{q}$, $q=1,2,...,n$, is the $q$-th received and equalized symbol and $b_{(q-1)M + j}$, $j=1,2,..,M$, is the $j$-th bit in the $q$-th equalized symbol. This definition of \acp{LLR} leads to the following bit error probability:
\begin{equation}\label{eq:bit_error}
v_l^{{(i)}} = \frac{1}{1 + e^{|\Lambda_{l}^{{(i)}}|}}\,,l=1,2,...,p\,,\,i=1,2.
\end{equation}
Against the background of \cite{early_harq_schemes2}, we provide in Eq.~(\ref{eq:bit_error}) a corrected version for the bit error probability \cite[Eq.~(7)]{early_harq_schemes2}. 

In order to reduce the high dimensionality of $p$ bit error estimates $v_l, l=1,...,p$, we average over all received bit error estimates:
\begin{equation}
    \overline{v}^{(i)} := \frac{1}{p} \sum_{l=1}^p v_l^{(i)}\,,i=1,2.
\end{equation}
We apply first a local logistic regression at each \ac{RRH} that is fed with $\overline{v}^{(i)}$ and the received \ac{SNR} $\gamma^{(i)}$. We train the local logistic regression with the \ac{BBU} decoding outcome as the ground truth. We apply $l_2$ regularization, see \cite{scikit_lr} for more details, and balanced weight classes to the logistic regression using the liblinear solver from the scikit-learn package \cite{scikit-learn}. The local feedback function $T_\mathrm{LLR}^{(i)}, i=1,2,$ is given as
\begin{equation}
    T_\mathrm{LLR}^{(i)} (\gamma^{(i)},\overline{v}^{(i)}) := Q\left(\frac{\exp(\theta_{20} + \begin{pmatrix}\gamma^{(i)}& \overline{v}^{(i)}\end{pmatrix} \boldsymbol{\theta}_2)}{1+\exp(\theta_{20} + \begin{pmatrix}\gamma^{(i)}&\overline{v}^{(i)}\end{pmatrix} \boldsymbol{\theta}_2)}\right),
\end{equation}
where $Q$ is the quantization function and $\{\theta_{20} \in \mathbb{R}, \boldsymbol{\theta}_2 \in \mathbb{R}^2\}$ is the parameter set of the logistic regression. The quantization function $Q$ is determined analogously to the \ac{Q-SNR} scheme. The range between the minimum and maximum value from the training set is divided into uniform intervals, where each interval assigns the value of its center. After having generated the local feedback $T_\mathrm{LLR}^{(i)}$, we again use a logistic regression at the \ac{UE} to learn the feedback combination rule $F$ \cite{hastie_09_elements-of.statistical-learning}:
\begin{equation}
    P_{\mathrm{ACK}|T_\mathrm{LLR}^{(1)} T_\mathrm{LLR}^{(2)}} \approx P_\mathrm{LLR} (T_\mathrm{LLR}^{(1)},T_\mathrm{LLR}^{(2)})
\end{equation}
with
\begin{equation}
    P_\mathrm{LLR} (T_\mathrm{LLR}^{(1)},T_\mathrm{LLR}^{(2)}) := \frac{\exp(\theta_{30} + \begin{pmatrix}T_\mathrm{LLR}^{(1)}&T_\mathrm{LLR}^{(2)}\end{pmatrix} \boldsymbol{\theta}_3)}{1+\exp(\theta_{30} + \begin{pmatrix}T_\mathrm{LLR}^{(1)}&T_\mathrm{LLR}^{(2)}\end{pmatrix} \boldsymbol{\theta}_3)}\,,
\end{equation}
where $\{\theta_{30} \in \mathbb{R}, \boldsymbol{\theta}_3 \in \mathbb{R}^2\}$ is the learnt parameter set of the logistic regression. Then, the combination rule is defined by $F_\mathrm{LR-SNR}(T_\mathrm{LLR}^{(1)},T_\mathrm{LLR}^{(2)}) := P_\mathrm{LLR}(T_\mathrm{LLR}^{(1)},T_\mathrm{LLR}^{(2)}) \stackunder{\stackon{\lessgtr}{\scriptstyle \mathrm{NACK}}}{\scriptstyle \mathrm{ACK}} C_\mathrm{LLR}$, where $C_\mathrm{LLR}$ is an appropriately chosen constant. This scheme is referred to as \ac{LR-LLR}.

\begin{figure}[t]
  \centering
  \ifone
  \includegraphics[width=.55\columnwidth]{distributed_lr}
  \else
  \ifarxiv
  \includegraphics[width=.9\columnwidth]{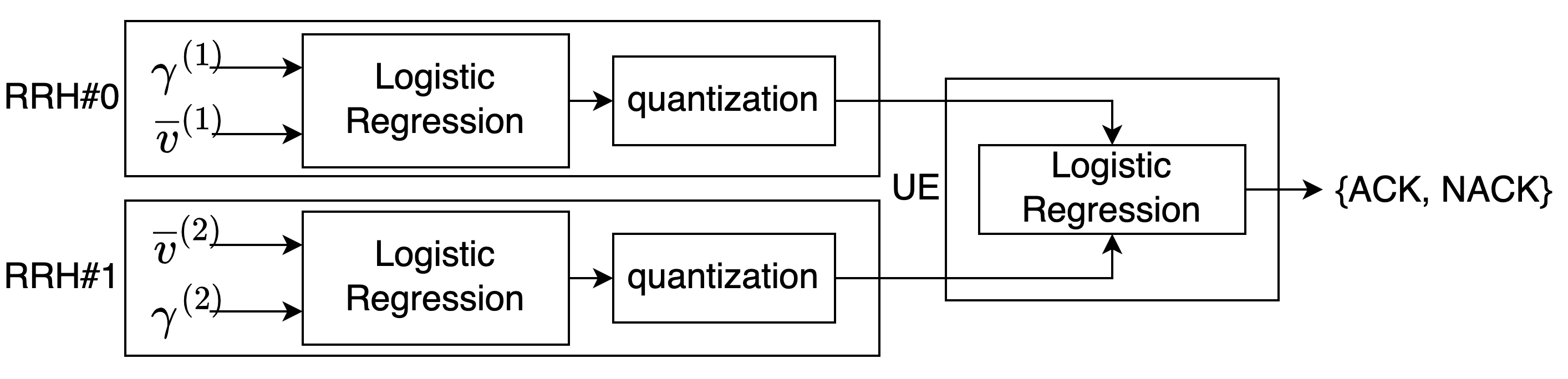}
  \else
  \includegraphics[width=.9\columnwidth]{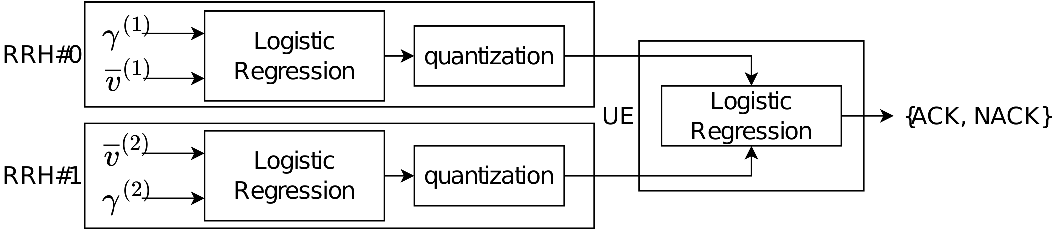}
  \fi
  \fi
  \caption{\ac{LR} based distributed prediction approach.}
  \label{fig:distributed_lr}
\end{figure}

\subsection{Subcode-based HARQ prediction}
As \ac{LLR}-based schemes, subcode-based \ac{HARQ} predictors take the \acp{LLR} as a basis. However, instead of assuming i.i.d. components of the signal vector $X$, the subcode-based prediction considers constraints defined by the parity-check matrix $\boldsymbol{P} \in \mathbb{F}_2^{L \times nM}$. The relation between the bit vector $\boldsymbol{b} := (b_{i,j}) \in \mathbb{F}_2^{nM}$, $i=1,2,...,n$, $j=1,2,...,M$, and the parity check matrix is defined as $ \boldsymbol{P}\boldsymbol{b}^\mathrm{T} = \boldsymbol{0}$.
Message passing decoders, such as the min-sum implementation, iteratively update the \acp{LLR} based on $\boldsymbol{P}$, which is described by:
\begin{equation}
\label{eq:ap_llr}
\Lambda_{l,k}^{{(i)}} = \Lambda_{l,k-1}^{{(i)}}  + \sum_{m \in \mathcal{M}(l)} \delta_{m,k}^{{(i)}} \,,\,i=1,2\,,
\end{equation}
where $\mathcal{M}(l)$ is the set of check nodes that are associated with the bit $b_{l}$, $\delta^{(k)}_{m,k}$ is the check node to variable node message at the $k$-th iteration, and $\Lambda_{l,k}^{(i)}$ is the updated \ac{LLR} at the $k$-th iteration with $\Lambda_{l,0}^{(i)} := \Lambda_{l}^{(i)}$. In contrast to tree codes where message passing decoders always converge to the best solution, for modern \ac{LDPC} codes, the evolution of the \acp{LLR} can be interpreted as a sequence which may or may not converge to a ”degraded" marginalization \cite{10.5555/971143}. Compared to considering only the received \acp{LLR}, this behavior provides additional information on the healthiness of the received codeword.

\subsubsection{Supervised dual autoencoding 2-stage gaussian mixture model (DA2SGMM) for anomaly detection}\label{sec:dida_overview}
In the machine learning literature, autoencoders are well established for unsupervised anomaly detection tasks \cite{Zhou2021FeatureEW,zong2018dagmm,Alad} due to their unprecedented dimensionality reduction capabilities \cite{GoodBengCour16}. Also for \ac{HARQ} prediction purposes, a supervised autoencoder proposed in \cite{journal_eharq_paper} outplayed other machine learning techniques, such as logistic regression, random forests, and others. However, the approach in \cite{journal_eharq_paper}, which builds on the DAGMM architecture that was proposed for anomaly detection in \cite{zong2018dagmm}, assumes a scenario with a single receive point. In this section, we extend this autoencoder to handle two separated \acp{RRH}. 
This novel approach, referred to as \ac{DIDA}, achieves a dimensionality reduction of the input features. Furthermore, we use two independent classifiers at each \ac{RRH} to convert the compressed subcode features together with the received \ac{SNR} features to a decodability feedback, which is afterwards combined at the \ac{UE} classifier. 
\begin{figure}[t]
  \centering
  \ifone
  \includegraphics[width=.45\columnwidth]{distributed_sae}
  \else
  \ifarxiv
  \includegraphics[width=.9\columnwidth]{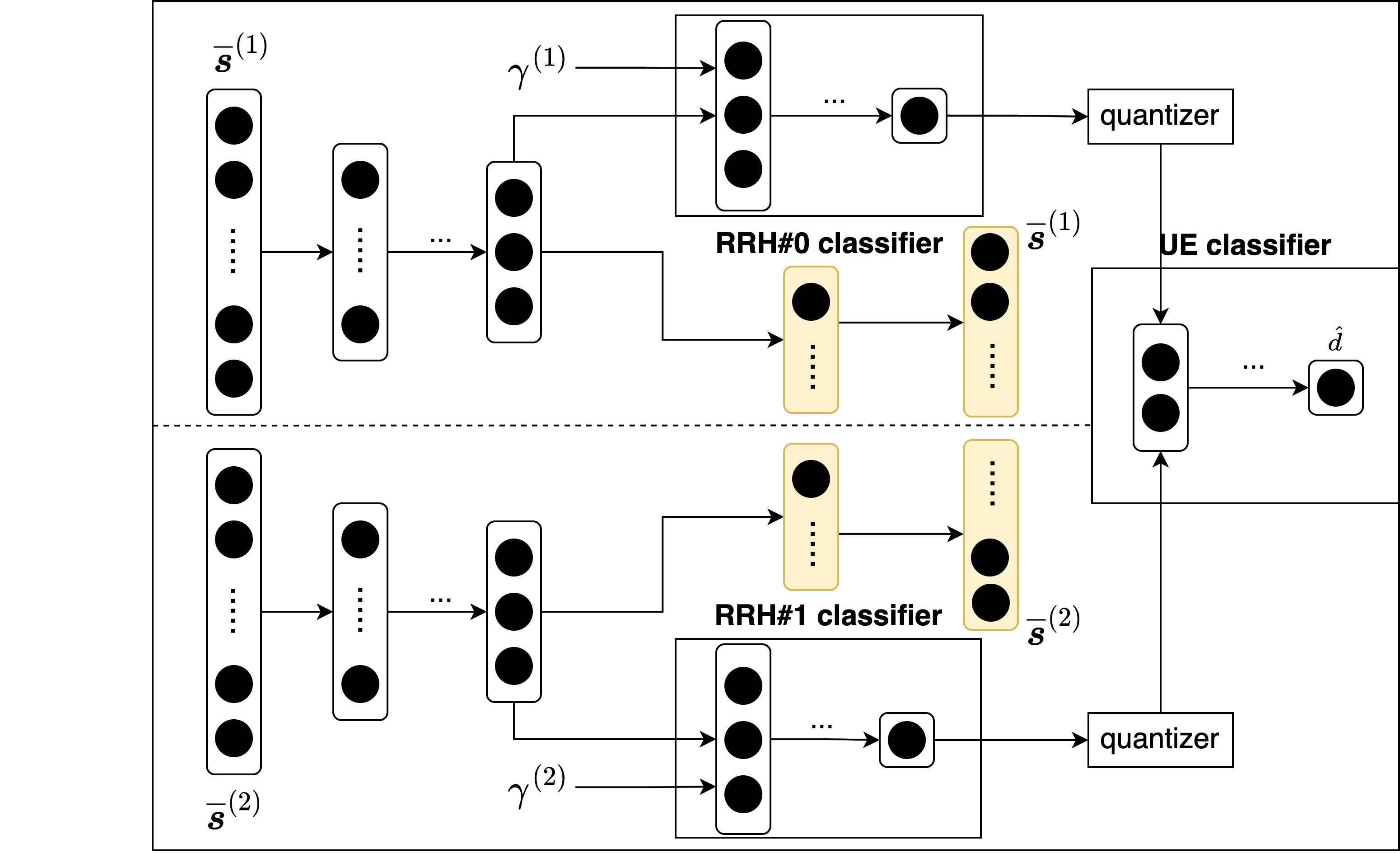}
  \else
  \includegraphics[width=.9\columnwidth]{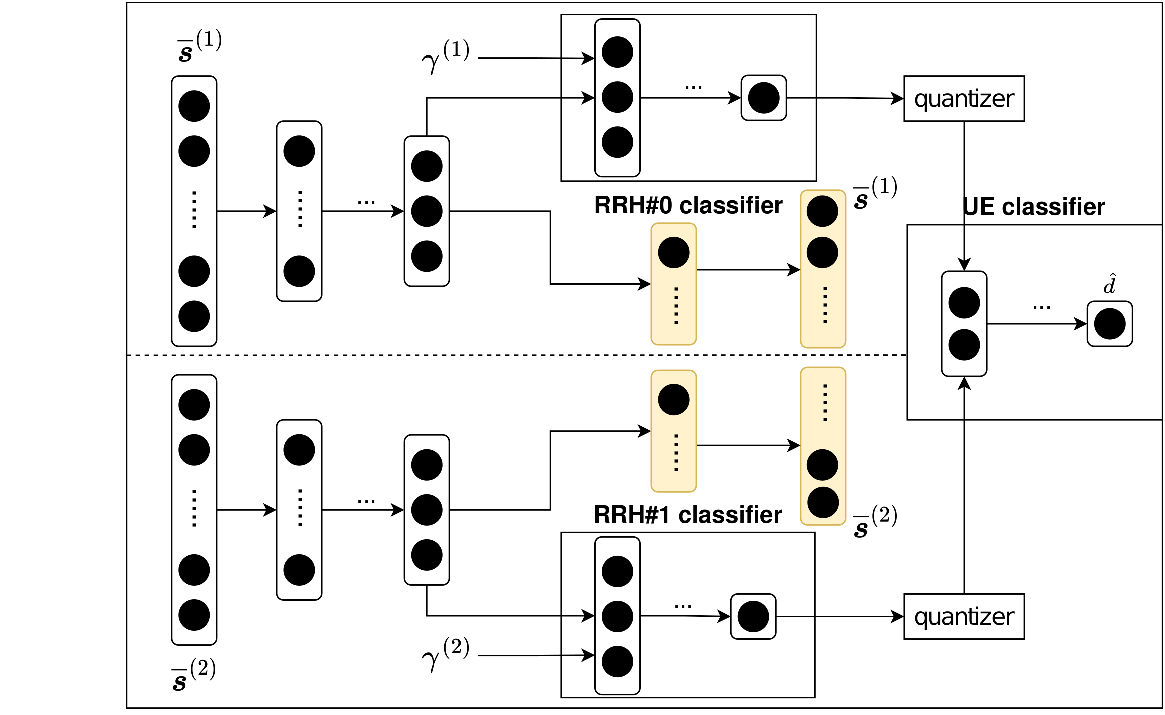}
  \fi
  \fi
  \caption{A supervised \ac{DIDA} to compress the feedback information at the RRHs and evaluate the combined result at the UE. The yellow boxes are only used for training purposes and are removed for inference.}
  \label{fig:distributed_sae}
\end{figure}
In Fig.~\ref{fig:distributed_sae}, we show the schematic design of the proposed \ac{DIDA} architecture. As can be seen, we incorporate the constraints of the architecture of the communication system directly into the setup of the \ac{DIDA}. The upper box represents the parts executed at the first \ac{RRH}, the lower box the parts at the second \ac{RRH} and finally, the right small box represents the network at the \ac{UE}.
We provide further details of the \ac{DIDA} setup and training in App.~\ref{sec:dida_details}.\\
The subcode features are generated from a partial decoding process at the \acp{RRH}. In particular, the subcode features at the \acp{RRH} are represented as:
\begin{equation}
    \overline{\boldsymbol{s}}^{(i)} := \begin{pmatrix} \overline{s}_1^{(i)}&...&\overline{s}_{N_\mathrm{sc}/4}^{(i)} \end{pmatrix}\,,i=1,2,
\end{equation}
where
\begin{equation}\label{eq:mean_sc}
    \overline{s}_k^{(i)} :=  \frac{1}{4p}\sum_{j=1}^4 \sum_{l=1}^p \frac{1}{1 + e^{|\Lambda_{l,4(k-1)+j}^{(i)}|}}\,,\,k=1,...,N_\mathrm{sc}/4.
\end{equation}

In contrast to the previous schemes, the training is performed in an end-to-end manner. Hence, we do not have to distinguish the feedback $T^{(i)}$ and the combination rule $F$. Instead, we can see the whole network, represented by $P_\mathrm{DA2SGMM}$, as part of the combination rule $F_\mathrm{DA2SGMM}\left(\overline{\boldsymbol{s}}^{(1)}, \gamma^{(1)}, \overline{\boldsymbol{s}}^{(2)}, \gamma^{(2)}\right) := P_\mathrm{DA2SGMM} \stackunder{\stackon{\lessgtr}{\scriptstyle \mathrm{NACK}}}{\scriptstyle \mathrm{ACK}} C_\mathrm{DA2SGMM}$, where $C_\mathrm{DA2SGMM}$ again is a constant controlling the trade-off between false-positives and false-negatives.

\subsection{Complexity comparison}\label{sec:complexity}


\begin{table*}[t]
\centering
\caption{Memory consumption in number of stored floating point variables and computational complexity in terms of elementary floating point operations.}
\label{tab:mem_comp_consm}
\begin{tabular}{|l|l|l|l|l|}
\hline
Scheme&Memory consumption& Memory consumption&Computational complexity&Computational complexity\\
&at the \ac{UE}&at each \ac{RRH}&at the \ac{UE}&at each \ac{RRH}\\
\hline
\ac{Q-SNR}&1&1&$C_\mathrm{+,*}$&$C_\mathrm{quant}$\\
\hline
\ac{LR-LLR} &$5$ & $5$ & $5 C_\mathrm{+,*} + C_\mathrm{/} + C_\mathrm{exp}$ & $5 C_\mathrm{+,*} + C_\mathrm{/} + C_\mathrm{exp} + C_\mathrm{quant}$\\
\hline
\ac{DIDA} & $407$ & $3841$ &$672 C_\mathrm{+,*} + C_\mathrm{/} + C_\mathrm{exp}$ &$7328 C_\mathrm{+,*} + C_\mathrm{/} + C_\mathrm{exp} + C_\mathrm{quant}$\\
\hline
\end{tabular}
\end{table*}

The different \ac{HARQ} prediction strategies come at different costs in terms of computations and memory. In particular, the required processing time, which results from the computational complexity, is critical for low-latency applications. In order to compare the different schemes, we assume that the \ac{SNR} and \ac{LLR} features themselves are available without any processing cost. In \cite{decoding_latency}, the decoding latency of a flexible offset min-sum \ac{LDPC} decoder is given by
\begin{equation}
    \delta_\mathrm{dec} = \left\lceil \frac{N d_v}{Z} \right\rceil \frac{I}{f}\,,
\end{equation}
where $N$ is the size of the codeword, $d_v$ is the average variable node degree, $Z$ is the lifting size of the code, i.e. 104, $I$ is the number of performed iterations and $f$ is the clock rate of the decoder. This decoder type is implementation-wise very similar to the optimized min-sum \ac{LDPC} algorithm, which has been used for the simulations. Applying to the used subcode and assuming a decoder frequency of \unit[1]{GHz}, as motivated in \cite{decoding_latency}, we obtain a decoding latency of \unit[305]{ns} for the partial decoding to obtain the subcode features from the \acp{LLR}.

For the evaluation of the classifiers, we determine the amount of memory required to store the model parameters and input features and the number of elementary floating-point operations. Furthermore, to validate the estimated we perform a processing time measurement of a single-threaded implementation of the schemes on an Intel(R) Xeon(R) CPU E5-2687W v3 @ 3.10GHz processor. The \ac{LR-LLR} scheme uses logistic regressions with 2 input features each. Hence, besides the features themselves, the parameter set of the logistic regression also has to be stored on the devices. This results to a memory consumption of $C_{\mathrm{mem},\mathrm{LR}} = 5$. Furthermore, the computational complexity is given by $C_{\mathrm{comp},\mathrm{LR}} = 5 C_\mathrm{+,*} + 1 C_\mathrm{/} + 1 C_\mathrm{exp}$, where $C_\mathrm{+,*}$ is the computational complexity of an elementary multiplication or addition, $C_\mathrm{/}$ is the computational complexity of an elementary division and $C_\mathrm{exp}$ is the complexity of computing the exponential function. We use $C_\mathrm{quant}$ to designate the computational complexity of the quantization. Different from the logistic regression, the \ac{DIDA} scheme is built up by multiple \ac{FC} layers, see App.~\ref{sec:dida_details} for more details. The overall memory consumption of an \ac{FC} layer results to
\begin{align}
C_{\mathrm{mem},\mathrm{FC}} = (N_\mathrm{in} + 3) N_\mathrm{out}\,.
\end{align}
Furthermore, the overall computational complexity of an FC layer is given by
\begin{align}
C_{\mathrm{comp},\mathrm{FC}} = 2 N_\mathrm{out} (N_\mathrm{in} + 1) C_\mathrm{+,*}\,.
\end{align}
The Softmax layer does not have any stored parameters and hence, $C_{\mathrm{mem},\mathrm{SM}} = 0$. The computational complexity is given by $C_{\mathrm{comp},\mathrm{SM}} = 2 C_\mathrm{+,*} + 1 C_\mathrm{/} + 1 C_\mathrm{exp}$.

\begin{table*}[t]
\centering
\caption{Single-threaded processing time on different processor platforms.}
\label{tab:complex}
\begin{tabular}{|l|l|l|l||l|}
\hline
Scheme&Device&Number of&Estimated processing time& Actual processing time\\
&&operations&Raspberry Pi 3 Model B&Intel(R) Xeon(R) CPU E5-2687W v3 @ 3.10GHz\\
&&&approx. 4~GFLOPS \cite{BASFORD2020278}&approx. 49.6~GFLOPS, obtained from linpack\\
\hline
\ac{Q-SNR} &\ac{UE}&1&\unit[0.3]{ns}&\unit[0.4]{ns}\\
\hline
\ac{Q-SNR} &\ac{RRH}&4&\unit[1.0]{ns}&\unit[24.4]{ns}\\
\hline
\ac{LR-LLR} &\ac{UE}&17&\unit[4.3]{ns}&\unit[72.8]{ns}\\\hline
\ac{LR-LLR} &\ac{RRH}&21&\unit[7.0]{ns}&\unit[158.5]{ns} \\\hline
\ac{DIDA} &\ac{UE}&692&\unit[173.0]{ns}&\unit[102.1]{ns} \\\hline
\ac{DIDA} &\ac{RRH}&7351&\unit[1837.8]{ns}&\unit[519.0]{ns}\\\hline
\end{tabular}
\end{table*}

\begin{table}[t]
\centering
\caption{Estimated parallelized processing time of DA2SGMM on a Raspberry Pi 3 platform.}
\label{tab:complex_par}
\begin{tabular}{|l|l|l|l|l|}
\hline
Number&Device&Number of&Scaling&Estimated processing time\\
of cores&&operations&factor&Raspberry Pi 3 Model B\\
&&per core&&approx. 4~GFLOPS \\
&&&&(per core) \cite{BASFORD2020278}\\
\hline
8&\ac{UE}&103&6.72&\unit[25.8]{ns}\\\hline
8&\ac{RRH}&949&7.75&\unit[237.3]{ns}\\\hline
16&\ac{UE}&85&8.14&\unit[21.3]{ns}\\\hline
16&\ac{RRH}&536&13.71&\unit[134.0]{ns}\\\hline
\end{tabular}
\end{table}

Tab.~\ref{tab:mem_comp_consm} summarizes the overall complexity of all schemes. Obviously, the \ac{Q-SNR} scheme has the least memory consumption as well as the least computational complexity at all devices. Compared to that, the \ac{LR-LLR} slightly increases the memory consumption and computational complexity at the \acp{RRH}. The \ac{DIDA} clearly has the highest memory consumption and computational complexity on all devices. The complexity of the elementary operations can be approximated by weights corresponding to the number of performed floating point operations: $C_\mathrm{+,*} := 1$, $C_\mathrm{/} = 4$ and $C_\mathrm{exp} = 8$ \cite{osti_6574703}. For the quantization, we take PyTorch's FakeQuantize implementation as baseline \cite{fakequant_pytorch}. Hence, the computational complexity of the quantization results to $C_\mathrm{quant} = 4 C_\mathrm{+,*}$. Tab.~\ref{tab:complex} shows the estimated processing times for a Raspberry Pi 3 processor and the results of actual processing time measurements on an Intel Xeon processor. In a practical implementation, the processing time may vary based on the capabilities of the processor platform, the latency of memory access, the efficiency of the implementation and many more factors. However, this impact also heavily depends on the actual implementation. This also explains discrepancies between the estimated processing times and the actual processing times. Obviously, the quantization operation requires much more time on the Intel processor than estimated. In contrast to the \ac{DIDA} that uses PyTorch also for the quantization, we use scikit-learn's KBinsDiscretizer for the quantization of the other schemes. Also, the implementation of the linear regression is not optimized for the particular case and hence, performs unnecessary double calculations. Furthermore, we do not consider memory access delays for our estimated processing times. However, with specialized hardware, such as GPUs or even TPUs, a significantly better performance is to be expected. In Tab.~\ref{tab:complex_par}, we show a simple estimation of the processing times of the \ac{DIDA} on multi-core platforms. We assume that matrix multiplications resulting in an output vector, i.e. a linear layer, can easily be parallelized by computing each entry of the output vector separately. We observe that the processing time on the \ac{RRH} scales almost linearly up to 16 cores. In contrast to that, on the \ac{UE}, 16 cores only have a very small advantage compared to 8 cores. However, a \ac{UE} processing time of \unit[25.8]{ns} already is very small. Nevertheless, the processing times are at most in the small \textmu s range on single-threaded platforms, which is extremely small even compared to stringent latency budgets, such as \unit[1]{ms} and even \unit[0.1]{ms}. In particular, specialized hardware, such as GPUs and TPUs, are widely used to run neural networks. Overall, our evaluation shows that the processing time are expected to be sufficiently small on practical platforms.

\subsection{Data transmission model}

\begin{figure}[t]
  \centering
  \ifone
  \includegraphics[width=.55\columnwidth]{proactive_harq}
  \else
  \ifarxiv
  \includegraphics[width=.9\columnwidth]{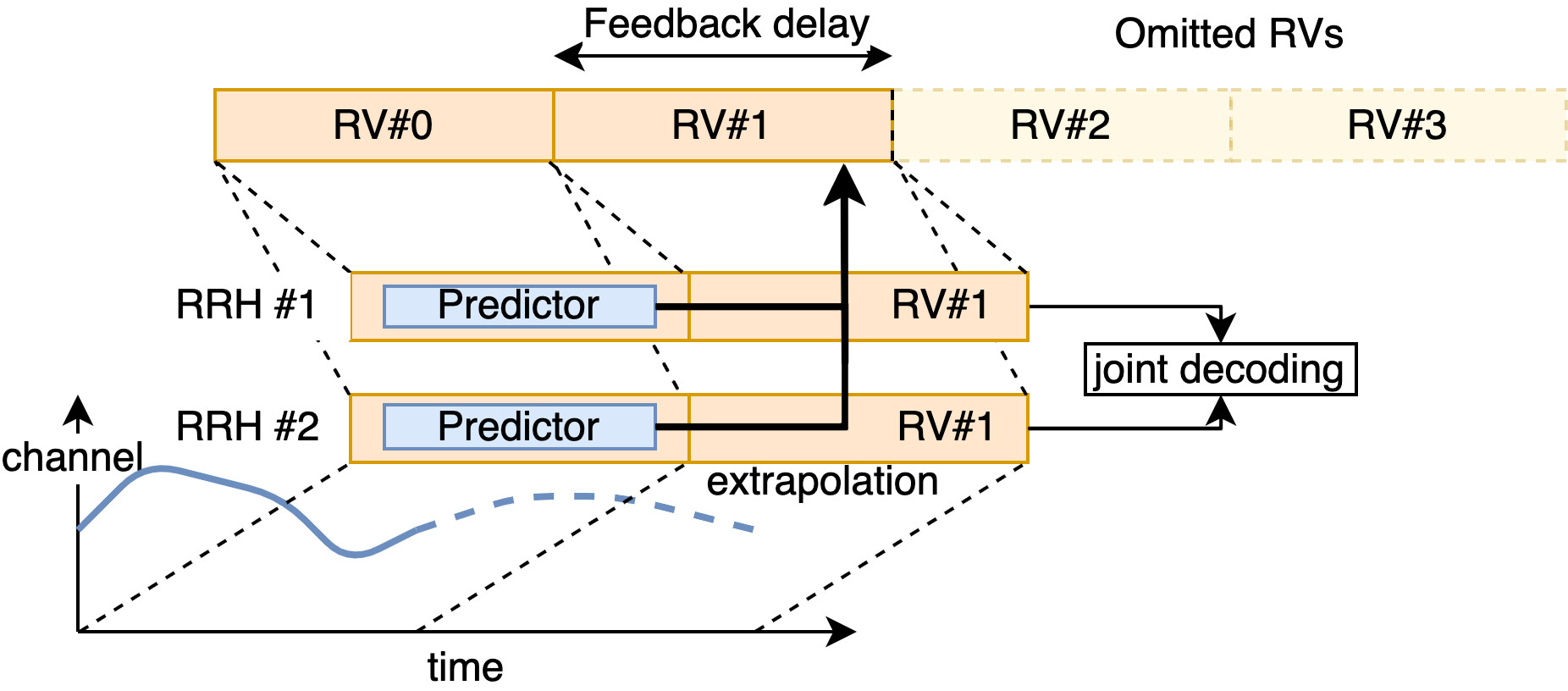}
  \else
  \includegraphics[width=.9\columnwidth]{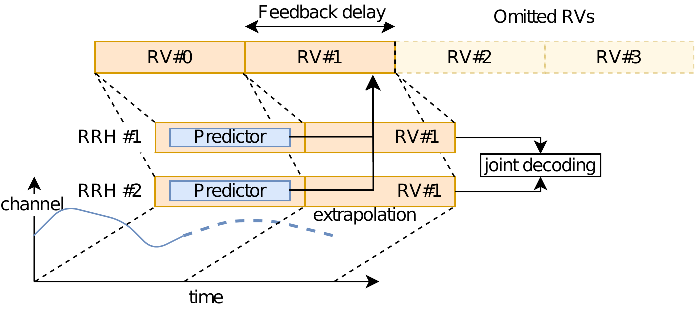}
  \fi
  \fi
  \caption{Data transmission model incorporating the feedback delay.}
  \label{fig:proactive_harq}
\end{figure}
We assume an incremental redundancy \ac{HARQ} protocol with up to four \acp{RV}. In our simulation setup, an \ac{RV} spans over 14 \ac{OFDM} symbols in time, which is equivalent to \unit[15.63]{\textmu s}.
The \ac{UE} transmits the \acp{RV} in a consecutive manner, as depicted in Fig.~\ref{fig:proactive_harq}. After receiving each \ac{RV}, the \acp{RRH} provide feedback using the previously described prediction schemes. The \ac{UE} decides based on a combination rule $F$ whether further \acp{RV} are required or not. This approach enables a good trade-off between reliability and spectral efficiency since some \acp{RV} are omitted, if an early decoding is successfully predicted. After having received all \acp{RV} from the \ac{UE}, the \acp{RRH} forward the received signals to the \ac{BBU} where a single decoding attempt is conducted. 
In this work, we simulate four \acp{RV}; hence, there are three prediction points:
\begin{enumerate}
    \item the \textbf{first prediction point (Pos\#1)}, which uses the received \ac{RV}\#0 to decide whether \ac{RV}\#2 is required or not,
    \item the \textbf{second prediction point (Pos\#2)}, which uses \ac{RV}\#0 and \ac{RV}\#1 to decide whether \ac{RV}\#3 is required or not.
    \item the \textbf{blockage prediction point (Pos\#3)}, which uses \ac{RV}\#0-2 to decide whether blockage is detected or not.
\end{enumerate}
A positive prediction at the first prediction point causes the \ac{UE} to stop transmitting further \acp{RV}. Hence, the second prediction point would not be reached in that case. However, this depends on the particular prediction scheme and modeling this accurately would require to incorporate the prediction directly into the link-level simulations, which increases the complexity extremely. Instead, we assume that all prediction schemes correctly identify transmissions that are already decodable with the first \ac{RV} and only the remaining transmissions reach the second prediction point.

\subsection{Evaluation methodology}\label{sec:eval_method}
In addition to achieving the reliability and the latency targets which are mandatory requirements, the performance of the \ac{HARQ} prediction schemes can be compared in terms of their achieved throughput. In contrast to commonly used classifier performance metrics, such as precision and recall, the throughput provides a measure with direct takeover to practical scenarios. Especially, when considering edge cases with extremely low-reliability requirements, common classifier metrics may not provide a good metric for comparison \cite{journal_eharq_paper}.\\
Based on the renewal-reward theorem \cite{ross1996stochastic}, the throughput is expressed as $\eta = \frac{\mathbb{E}[\mathcal{R}]}{\mathbb{E}[L(T)]}$, where $\mathbb{E}[\mathcal{R}]$ is an expected reward and, $\mathbb{E}[L(T)]$ is the expected transmission latency with $L(T)$ as defined in (\ref{eq:lat_early}) and (\ref{eq:lat_harq}) and $T \in \mathbb{N}^+$ being the number of requested \acp{RV}. Furthermore, the reward is $\mathcal{R} := 0$ in case the transmission failed within the latency budget. In case the transmission was successful,  $\mathcal{R}$ is $N_\mathrm{packet}$, the size of the packet in nats. Hence, the expected reward is given as $\mathbb{E}[\mathcal{R}] := (1 - \epsilon_\mathrm{tot}) N_\mathrm{packet}$, where $\epsilon_\mathrm{tot}$ is the associated total error probability. The transmission latency of the prediction schemes is composed of multiple components:
\begin{equation}\label{eq:lat_early}
    L_\mathrm{e}(T) = \begin{cases}
        \!\begin{aligned}
         &T \delta_\mathrm{RV} + (T-1)\delta_{\mathrm{exc}}
         \end{aligned}&\text{if }1 \leq T \leq T_\mathrm{max}\\
         \!\begin{aligned}
         &T_\mathrm{max} \delta_\mathrm{RV} + L_\mathrm{blk}\\
         &+ (T_\mathrm{max}-1)\delta_{\mathrm{exc}}
         \end{aligned}&\text{if }T > T_\mathrm{max}
    \end{cases}\,,
\end{equation}
where $\delta_\mathrm{exc} := \max(\delta_\mathrm{proc} + \delta_\mathrm{fb} - \delta_\mathrm{RV}, 0)$ and $\delta_\mathrm{RV}$ is the time to transmit an \ac{RV}, $\delta_\mathrm{proc}$ is the processing time required by the specific prediction scheme and $\delta_\mathrm{fb}$ is the time required to transmit the feedback. Furthermore, $L_\mathrm{blk}$ is a latency penalty. For simplicity, we assume $L_\mathrm{blk} = T_\mathrm{blk} \delta_\mathrm{RV}$, where $T_\mathrm{blk}$ is a spectral efficiency penalty when blockage is detected. In contrast to that, the latency of a regular \ac{HARQ} system is composed as:
\begin{equation}\label{eq:lat_harq}
    L_\mathrm{r}(T) = \begin{cases}
         T \delta_\mathrm{RV} + (T-1)(\delta_\mathrm{fh} + \delta_\mathrm{fb})&\text{if }1 \leq T \leq T_\mathrm{max}\\
         \!\begin{aligned}
         &T_\mathrm{max} \delta_\mathrm{RV} + L_\mathrm{blk}\\
         &+ (T_\mathrm{max}-1)(\delta_\mathrm{fh} + \delta_\mathrm{fb})
         \end{aligned}&\text{if }T > T_\mathrm{max}
    \end{cases}\,,
\end{equation}
where $\delta_\mathrm{fh}$ is the fronthaul round-trip time, which is the time required for transporting the received signal vectors to the \ac{BBU}, decoding the packet at the \ac{BBU}, and sending the result back to the \acp{RRH}.\\
The probability distribution of $T$ for a feedback delay of $\delta := 1$ is determined by the performance of the different estimators:
\begin{multline}\label{eq:expected_transmissions}
    \mathbb{P}[T = t] = \begin{cases}
        \!\begin{aligned}
        &p_{t-2} \epsilon_{t-1}\alpha_{t-1}\\
        &+ p_{t-2}(1-\epsilon_{t-1})\\
        &\cdot (1-\beta_{t-1})
        \end{aligned}&\text{if }2 \leq t \leq T_\mathrm{max}\\
        p_{T_\mathrm{max}-1}&\text{if }t = T_\mathrm{max}+T_\mathrm{blk}\\
        0&\text{otherwise}
    \end{cases}
\end{multline}
with
\begin{equation}
    p_t := \prod_{j=1}^{t}\epsilon_j(1-\alpha_j) + (1-\epsilon_j)\beta_j\,,
\end{equation}
where $T_\mathrm{max}$ is the maximum number of transmissions, $T_\mathrm{blk}$ is an additional spectral efficiency penalty in case blockage is detected, $\epsilon_{j}$, $j \in \{1,2,...,T_\mathrm{max}-1\}$, are the error probabilities at the $(j+1)$-th \ac{RV} given that previous \acp{RV} were unsuccessful, $\alpha_{j}$, $j \in \{1,2,...,T_\mathrm{max}-1\}$, are the false positive probabilities, i.e. predicting an unsuccessful decoding as an ACK, and $\beta_{j}$, $j \in \{1,2,...,T_\mathrm{max}-1\}$, are the false negative probabilities, i.e. predicting a\deleted{n} successful decoding as a NACK. The false-positive and false-negative probabilities are defined as
\begin{equation}\label{eq:fp}
    \alpha_j := P_{D_{j+1}=\mathrm{NACK}|F_j=\mathrm{ACK}}\,,
\end{equation}
and
\begin{equation}\label{eq:fn}
    \beta_j := P_{D_{j+1}=\mathrm{ACK}|F_j=\mathrm{NACK}}\,,
\end{equation}
where $D_{j+1}$ is the decoding outcome with $(j+1)$ \acp{RV} and $F_j$ is the outcome of the $j$-th prediction.\\
The total error performance $\epsilon_\mathrm{tot}$ is determined by the error probabilities $\epsilon_i$ but also the false positive error probability $\alpha_i$. The total error performance in a non-blockage scenario is given by \cite{feedback_prediction_iiot}:
\begin{multline}\label{eq:total_error}
    \epsilon_{\mathrm{tot}|\mathrm{nb}} = \left(\sum_{t=1}^{T_\mathrm{max}-2} \left(\prod_{j=1}^{t-1} \epsilon_j(1-\alpha_j)\right) \epsilon_{t} \alpha_{t}\right)\\
    + \left(\prod_{i=1}^{T_\mathrm{max}-2} \epsilon_i (1-\alpha_i)\right) \epsilon_{T_\mathrm{max}-1}\,,
\end{multline}
where the false-positive, false-negative and error probabilities are estimated from non-blockage scenarios.
However, in a blockage scenario a sufficiently low error probability cannot be maintained and hence, alternative procedures, e.g. additional redundancy, switching to a lower frequency or adapting the beam, have to be initiated. However, instead of making an assumption on the specific blockage recovery scheme, we use the spectral efficiency penalty $T_\mathrm{blk}$ to model the blockage case. This also means that an effective blockage, i.e. non-decodability of the $T_\mathrm{max}$ \acp{RV}, has to be detected with the same target error probability. The probability of blockage misdetection is given by
\begin{multline}\label{eq:blockage_detection}
    \epsilon_{\mathrm{blk}|\mathrm{sb}} =  \left(\sum_{t=1}^{T_\mathrm{max}-3} \left(\prod_{j=1}^{t-1} \epsilon_j(1-\alpha_j)\right) \epsilon_t \alpha_t\right)\\
    + \left(\prod_{j=1}^{T_\mathrm{max}-3} \epsilon_j(1-\alpha_j)\right) \epsilon_{T_\mathrm{max}-1} \alpha_{T_\mathrm{max}-1}\,,
\end{multline}
where the false-positive, false-negative and error probabilities are derived from single-blockage scenarios.\\
The false-positive probabilities and the false-negative probabilities behave in a conflicting manner, where the trade-off between both can be controlled by adjusting the bias $s$ of the respective predictor. However, the functional relation between $\alpha_i(s)$ and $\beta_i(s)$, $i=1,...,T_\mathrm{max}-1$, is not known. Hence, we determine 1000 admissible pairs $(\alpha_i(s_j), \beta_i(s_j))$, $j=1,2,...,1000$, from the link-level simulations and interpolate them piece-wise linearly. Obviously, any false-positive false-negative curve of a reasonable predictor is a convex function, where at the extreme case when no prediction is possible, this function becomes a straight line connecting the points $(0,1)$ and $(1,0)$. Hence, a piece-wise linear interpolation is a conservative approximation, where the actual performance of the predictors at an interpolated point is expected to be better than the approximated value.

With $\boldsymbol{s} := (s_1, s_2, ..., s_{T_\mathrm{max}-1})$ being the vector of biases per prediction, we derive the following optimization problem for the expected number of \acp{RV} under no blockage:
\begin{equation}
\label{eq:prHARQ_opt}
    \begin{aligned}
        &\underset{\boldsymbol{s}}{\text{minimize}}&&\mathbb{E}[T(\boldsymbol{\alpha}_{\mathrm{nb}}(\boldsymbol{s}), \boldsymbol{\beta}_{\mathrm{nb}}(\boldsymbol{s}))]\\
        &\text{subject to} &&\epsilon_\mathrm{tot|nb}(\boldsymbol{\alpha}_{\mathrm{nb}}(\boldsymbol{s})) \leq \epsilon_{\mathrm{target}}\\
        &\text{and}
        &&\epsilon_\mathrm{blk|sb}(\boldsymbol{\alpha}_\mathrm{sb}(\boldsymbol{s})) \leq \epsilon_{\mathrm{target}}\,,
    \end{aligned}
\end{equation}
where $\epsilon_{\mathrm{target}}$ corresponds to the overall reliability requirement. We use the \ac{SLSQP} algorithm \cite{scikit_sqlsp} with a Monte-Carlo approach to numerically find a valid solution to the aforementioned optimization problem.

\subsection{Link-level simulation setup}\label{sec:lls}
\begin{table}[t]
\centering
\caption{Link-level simulation assumptions for training, hyper and test set generation.}
\label{tab:lls_assump}
\begin{tabular}{|l|l|}
\hline
\ac{TB} size& 1000 \\
in bits ($N_\mathrm{Bits}$)&\\
\hline
\ac{RV} length in bits& 2016\\
\hline
\ac{RV} duration ($\delta_{\mathrm{RV}}$)&14 OFDM symbols $\hat{=}$ 15.625~\textmu s\\
\hline
Transmission configuration& $T_\mathrm{max} = 4$, $\delta_\mathrm{fb} = 1$\\ 
\hline
Transmission bandwidth&40.0~MHz (6 PRBs)\\
\hline
Channel Code&\ac{5G} \ac{LDPC} (see \cite{5g_channel_coding_spec})\\
\hline
Modulation order and algorithm&4-QAM, $M=2$, Approximated LLR\\
\hline
Power allocation&Constant $E_b/N_0$\\
\hline
Waveform&3GPP OFDM, normal cyclic-prefix,\\
&960~kHz subcarrier spacing\\
\hline
\ac{SNR} range&no-blockage:\\
&\unit[4.0]{dB}, \unit[5.0]{dB}, \unit[6.0]{dB}, \unit[7.0]{dB}\\
&single-blockage:\\
&\unit[-7.2]{dB}, \unit[-6.2]{dB}, \unit[-5.2]{dB}, \unit[-4.2]{dB}\\ 
\hline
Target error rates $\epsilon_\mathrm{target}$&$1\cdot 10^{-4}$ (\unit[4]{dB}), $3.5\cdot 10^{-5}$ (\unit[5]{dB}),\\
&$1\cdot 10^{-5}$ (\unit[6]{dB}), $3.5\cdot 10^{-6}$ (\unit[7]{dB})\\
\hline
Channel type (no-blockage)&filtered \ac{CDL}-D 30~ns (Rician LOS),\\
&directional Rx/Tx antenna pattern,\\
&100~GHz, 50~km/h,\\
&ideal channel estimation\\
\hline
Channel type (single-blockage)&filtered \ac{CDL}-C 30~ns (Rayleigh LOS),\\
&directional Rx/Tx antenna pattern,\\
&100~GHz, 50~km/h,\\
&ideal channel estimation\\
\hline
Equalizer&Frequency domain MMSE\\
\hline
Decoder type&Min-Sum (50 iterations)\\
\hline
\end{tabular}
\label{tab:SIM}
\end{table}

To compare the performance of previously described \ac{HARQ} prediction schemes, we conduct link-level simulations to collect the required \ac{LLR}, subcode and channel estimation features. We choose the frame structure of the transmission, i.e. the mapping of the code block to resource elements and reference signals, e.g. \ac{DMRS}, in accordance with the Rel. 16 \ac{3GPP} specifications. Furthermore, we use a subcarrier spacing of \unit[960]{kHz}, which is currently being specified in the "NR operation up to 71 GHz" work item \cite{71_ghz_wid}. Tab.~\ref{tab:SIM} summarizes the link-level parameters. We assume that different \acp{UE} are scheduled on different orthogonal time-frequency resources and hence, interference from other transmitters can be neglected. In the sub-THz frequency spectrum, the use of beamforming is necessary due to the high pathloss, even for free space propagation. However, \ac{MIMO} schemes that allow dynamic beamforming are complex in the sense that they offer a large set of tunable parameters and transmission modes. Optimizing these goes beyond the scope of this work. Hence, we use a spatially filtered \ac{CDL-D} channel model that already incorporates the effects of beamforming \cite{3gpp_channel_models_100}, see Sec.~\ref{sec:lls} for more details.\\
We evaluate the performance in a \textbf{single-blockage} and a \textbf{no-blockage} scenario. We do not consider the case where both \acp{RRH} are blocked, as this would also make any communication at a feasible rate impossible. In the no-blockage scenario, we assume no \ac{SNR} difference between the two \acp{RRH}. In the single-blockage scenario, we assume the same \ac{SNR} at the non-blocked \ac{RRH} and an additional pathloss of \unit[11.2]{dB} at the blocked \ac{RRH}. This is the pathloss difference of a blocked and non-blocked channel with a \ac{UE} at \unit[20]{m} distance from both \acp{RRH} in the UMi scenario \cite{3gpp_channel_models_100}. At the \acp{RRH}, we use a spatially filtered \ac{CDL} channel model. In particular; at the un-blocked \ac{RRH}, we use \ac{CDL}-D, which is a \ac{LOS} channel model with Rician distributed \ac{LOS} component and Rayleigh distributed \ac{NLOS} components \cite{3gpp_channel_models_100}. At the blocked \ac{RRH}, we use the \ac{CDL}-C, which is a \ac{NLOS} channel model \cite{3gpp_channel_models_100}. Furthermore, we apply the directional antenna pattern, as specified in \cite{itu_antenna}, at both sides to generate a spatially filtered \ac{TDL} channel which models the effective channel between the \ac{UE} and the \acp{RRH}. The spatial filtration procedure is performed in accordance with \cite[Sec.~7.7.4]{3gpp_channel_models_100}. For the decoding of the received signal vector, we apply an optimized min-sum algorithm \cite{6174654} with 50 iterations. In contrast to that, the subcode prediction uses only 5 iterations. Hence, its complexity is only 1/10-th of the complexity of a full decoding attempt.

\section{Simulation Results}
In this section, we present the results for the different \ac{HARQ} prediction approaches. We train all schemes jointly on all \acp{SNR} except \unit[6]{dB}. The \ac{SNR} of \unit[6]{dB} is not used during training but only used for testing to evaluate the generalization performance of the models.

\subsection{False-positive and false-negative performance}
The false-positive and false-negative mispredictions determine the performance of the predictor, as we can see from Eq.~(\ref{eq:expected_transmissions}) and (\ref{eq:total_error}). Especially, the regime of low false-positives is of special interest because the cost of a false-positive misprediction is significantly higher than the cost of a false-negative misprediction. However, due to the finite size of the test sets, we have to deal with false-positives of zero, which makes a logarithmic scale unusable. Hence, a symmetrical logarithmic scale has been chosen for the false-positive axis. This scaling puts emphasis on the lower regime of the false-positive probabilities while keeping the zero point interpretable; however, special care has to be given to the linear scaling between the zero and the first step.

\begin{figure}
     \centering
        \ifarxiv
          \subfloat[]{\includegraphics[width=0.45\textwidth]{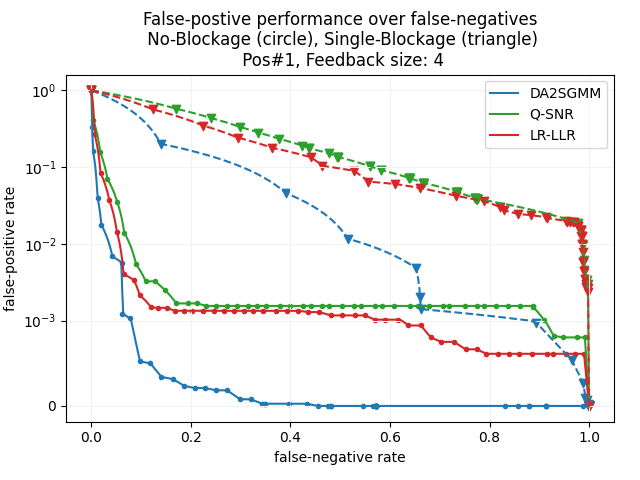}%
        \label{fig:fp_sc0_4b}}
        \hfil
        \subfloat[]{\includegraphics[width=0.45\textwidth]{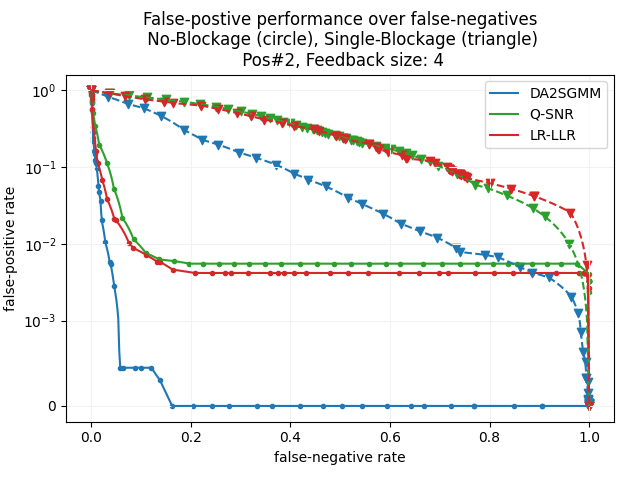}%
        \label{fig:fp_sc1_4b}}
          \fi
        
    \caption{False-positive prediction performance over false-negatives after the first RV and after second RV with 4 bit feedback. No-blockage (circle). Single-blockage (circle). (a) First prediction point. (b) Second prediction point.}
        \label{fig:fp_4b}
\end{figure}
Fig.~\ref{fig:fp_4b} shows the \ac{SNR}-averaged false-positive rate over the false-negative rate at the first and the second prediction points with a feedback size of 4 bits. In Fig.~\ref{fig:fp_sc0_4b}, we note that all schemes achieve a better performance in the no-blockage scenario compared to the single-blockage scenario. In particular, the \ac{Q-SNR} and \ac{LR-LLR} achieve in both scenarios a comparable performance, whereas the \ac{LR-LLR} performs slightly better than the other schemes except at very small false-negative rates in the no-blockage scenario. Furthermore, we note that the \ac{DIDA} clearly outperforms all other schemes in both scenarios. In the no-blockage scenario, it reaches zero mispredictions on the test set already at an false-negative rate of approximately $0.5$. In contrast to that, the other schemes reach zero mispredictions only at a false-negative rate of $1$. This behavior even reinforces at the second prediction point, seen in Fig.~\ref{fig:fp_sc1_4b}. Here, in the no-blockage scenario, the \ac{DIDA} reaches zero mispredictions already below a false-negative rate of $0.2$. In the single-blockage scenario, the performance of the \ac{DIDA} degrades slightly compared to the first prediction point. However, the performance of the other prediction schemes degrades significantly in both scenarios compared to the first prediction point.

\begin{figure}
     \centering
    \ifarxiv
           \subfloat[]{\includegraphics[width=0.45\textwidth]{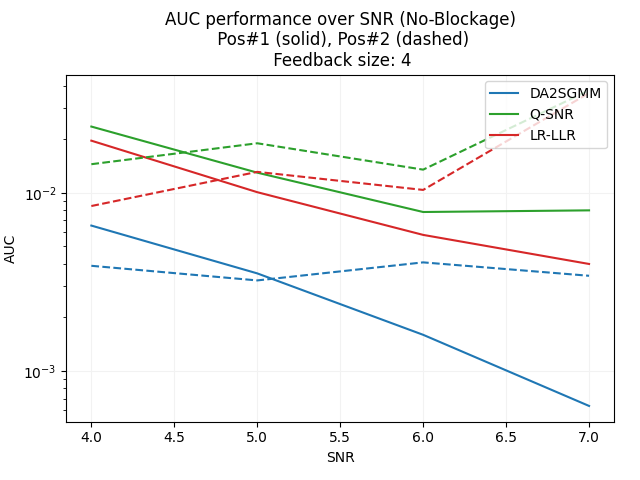}%
        \label{fig:auc_snr_nb}}
        \hfil
        \subfloat[]{\includegraphics[width=0.45\textwidth]{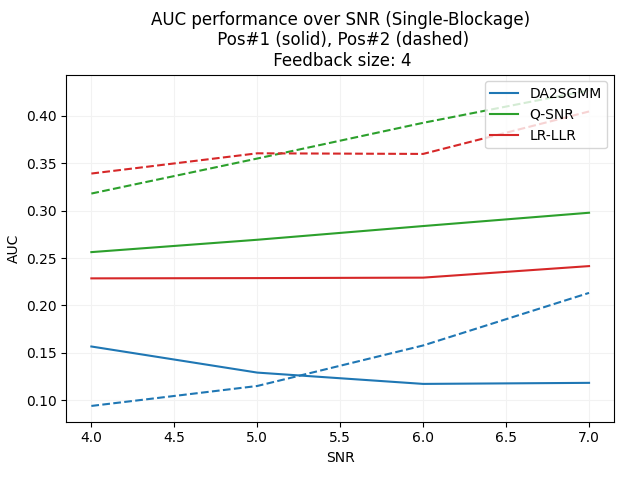}%
        \label{fig:auc_snr_sb}}
          \fi
    \caption{AUC performance over SNR with a feedback size of 4 bits. First prediction point (solid). Second prediction point (dashed). (a) No-blockage scenario. (b) Single-blockage scenario.}
        \label{fig:auc_snr}
\end{figure}
The false-positive-false-negative curve only shows the averaged performance. Hence, we also want to compare the performance of the schemes at specific \acp{SNR}. In particular, the \ac{SNR} of \unit[6]{dB} is of uttermost interest as this data was excluded from the training. Hence, we further introduce the notion of the \ac{AUC} as $AUC := \int_0^1 \tilde{f}_{\boldsymbol{\alpha},\boldsymbol{\beta}}(x)dx$, where $\tilde{f}_{\boldsymbol{\alpha},\boldsymbol{\beta}}(x)$ is the piece-wise linear interpolation of the false-positive-false-negative pairs $(\boldsymbol{\alpha},\boldsymbol{\beta})$. In Fig.~\ref{fig:auc_snr}, we present the \ac{AUC} performance in the no-blockage and single-blockage scenarios with a feedback size of 4 bits. In the no-blockage scenario, in Fig.~\ref{fig:auc_snr_nb}, we observe at the first prediction point that the \ac{AUC} decreases with increasing \ac{SNR}, i.e. the prediction accuracy increases. In particular, for the \ac{SNR} of \unit[6]{dB}, we note that none of the schemes show a particularly degraded \ac{AUC} performance. For the second prediction point, we observe that the \ac{AUC} tends to slightly increase with increasing \ac{SNR} for all schemes except the \ac{DIDA}, which shows an almost flat behavior over the \ac{SNR} range. As seen already in Fig.~\ref{fig:fp_4b}, it can be clearly seen that the \ac{DIDA} achieves by far the lowest \ac{AUC} at all \acp{SNR} and both prediction points. In the single-blockage scenario, in Fig.~\ref{fig:auc_snr_sb}, we observe the \ac{AUC} tending to increase with the \ac{SNR} except for the \ac{DIDA} at the first prediction point. Again, the \ac{DIDA} clearly outperforms the other schemes at all \acp{SNR} and both prediction points. Furthermore, we note as in the no-blockage scenario that no degradation of the \ac{AUC} at an \ac{SNR} of \unit[6]{dB} can be observed. Hence, a good generalization of all models may be assumed.

\subsubsection{Impact of feedback size}
In the previous section, we show results for a feedback size of 4 bits. However, the question of how many feedback bits are required is important for the practicability of the \ac{HARQ} prediction schemes, as more bits result in a significantly higher control signaling overhead.
\begin{figure}
     \centering
    \ifarxiv
           \subfloat[]{\includegraphics[width=0.45\textwidth]{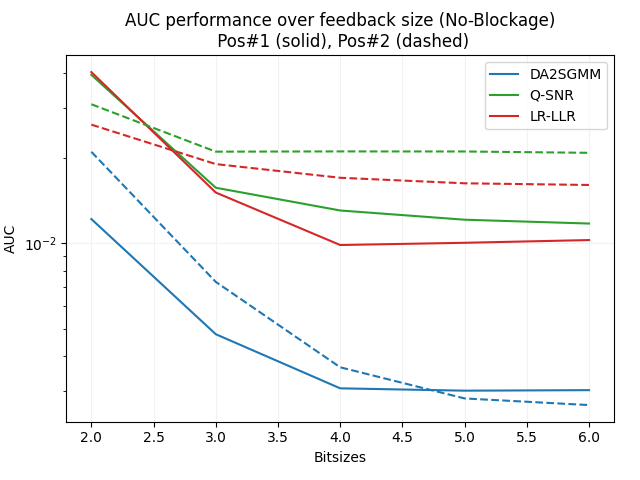}%
        \label{fig:auc_bit_nb}}
        \hfil
        \subfloat[]{\includegraphics[width=0.45\textwidth]{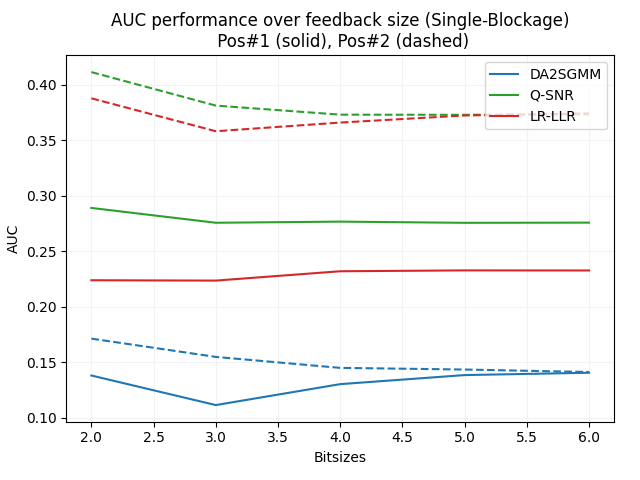}%
        \label{fig:auc_bit_sb}}
          \fi
    \caption{Averaged AUC performance over different feedback sizes. First prediction point (solid). Second prediction point (dashed). (a) No-blockage scenario. (b) Single-blockage scenario.}
        \label{fig:auc_bit}
\end{figure}
In Fig.~\ref{fig:auc_bit}, we show the \ac{SNR}-averaged \ac{AUC} over different feedback sizes in the no-blockage and the single-blockage scenarios, in Fig.~\ref{fig:auc_bit_nb} and Fig.~\ref{fig:auc_bit_sb}, respectively. In the no-blockage scenario, Fig.~\ref{fig:auc_bit_nb}, we clearly note a trend of lower \ac{AUC} at higher feedback sizes. This matches the intuition that more accurate feedback benefits the prediction accuracy. However, in the single-blockage scenario, in Fig.~\ref{fig:auc_bit_sb}, we observe a lower \ac{AUC} at lower feedback sizes. Although this seems counter-intuitive, this behavior is explained by the trade-off between the no-blockage \ac{AUC} and the single-blockage \ac{AUC}. Depending on the hyperparameters, the feedback size itself and in particular the \ac{ACK} weight class for the \ac{DIDA}, see Sec.~\ref{sec:dida_details}, the schemes train for a different trade-off at the different feedback sizes. Besides that, we observe that the \ac{DIDA} achieves the lowest \ac{AUC} at both prediction points and in both scenarios even compared to higher feedback sizes of the other schemes. Furthermore, we can see that all schemes profit from more feedback bits. In particular, we observe that \ac{Q-SNR} gains the most until 3 bits and only benefits slightly from more bits. The \ac{LR-LLR} and \ac{DIDA} mostly improve in terms of \ac{AUC} until 4 bits. Although, the \ac{DIDA} has an outlier for the first prediction point at 3 bits in the blockage scenario, as seen in Fig.~\ref{fig:auc_bit_sb}. 

\subsubsection{Blockage Detection}
In addition to the first and second \ac{HARQ} prediction points, the blockage prediction also plays a crucial role for practical applications.
\begin{figure}
         \centering
        \ifarxiv
           \subfloat[]{\includegraphics[width=0.45\textwidth]{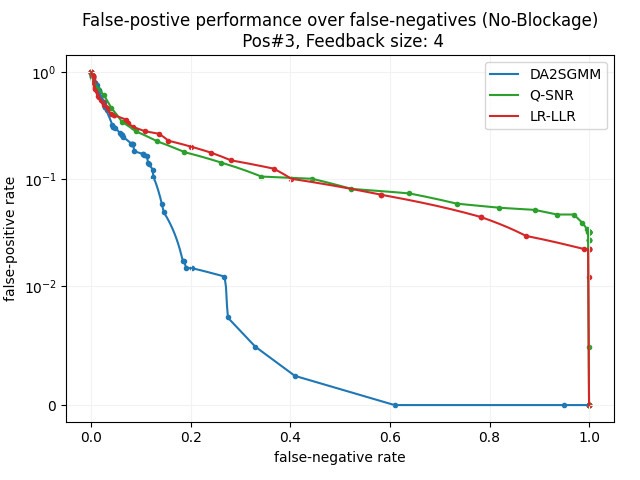}%
        \label{fig:fp_sc2_4b_nb}}
        \hfil
        \subfloat[]{\includegraphics[width=0.45\textwidth]{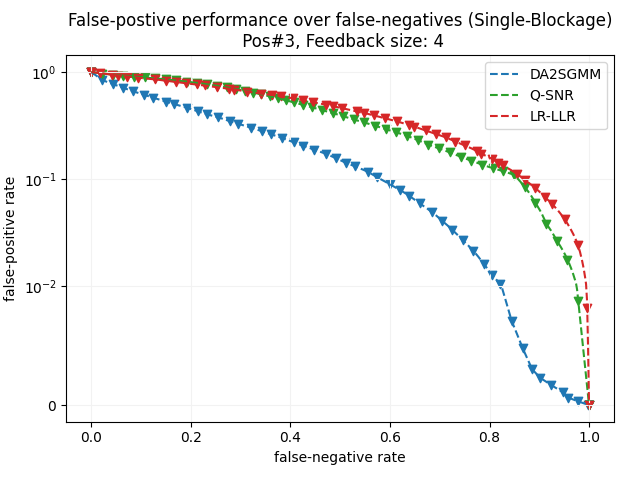}%
        \label{fig:fp_sc2_4b_sb}}
          \fi
    \caption{Blockage prediction at the third prediction point with a feedback size of 4~bits. (a) No-blockage scenario. (b) Single-blockage scenario.}
        \label{fig:fp_sc2_4b}
\end{figure}
In Fig.~\ref{fig:fp_sc2_4b}, we show the \ac{SNR}-averaged false-positive rate over the false-negative rate in the no-blockage and the single-blockage scenarios, in Fig.~\ref{fig:fp_sc2_4b_nb} and Fig.~\ref{fig:fp_sc2_4b_sb}, respectively. Similar to the previous prediction points, we observe generally a better performance for all schemes in the no-blockage scenario compared to the single-blockage scenario. Again, the \ac{DIDA} clearly achieves a significantly lower false-positive rate at the same false-negative rates. These results indicate a superior performance for the \ac{DIDA} scheme compared to the other schemes in terms of \ac{HARQ} prediction and also blockage detection.

\subsection{HARQ system performance}
In the previous section, we evaluated the false-positive and false-negative performance. However, the performance in a practical setup has to be shown to prove the efficiency of a scheme. Hence, we evaluate the different prediction schemes using the evaluation methodology described in Sec.~\ref{sec:eval_method}. 

\begin{figure}
     \centering
    \ifarxiv
        \subfloat[]{\includegraphics[width=0.45\textwidth]{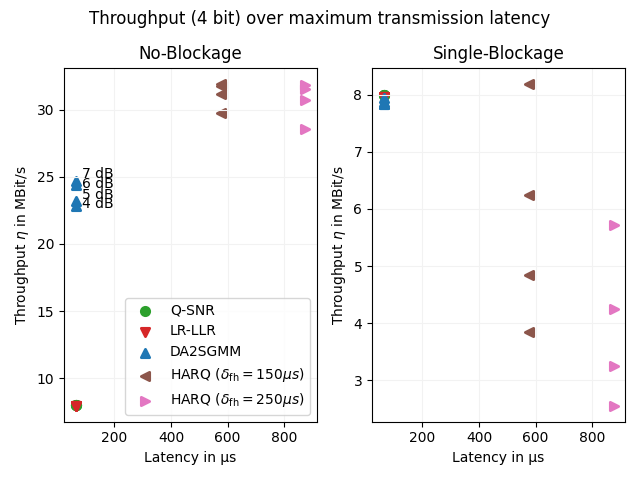}%
        \label{fig:harq_perf_0}}
        \hfil
        \subfloat[]{\includegraphics[width=0.45\textwidth]{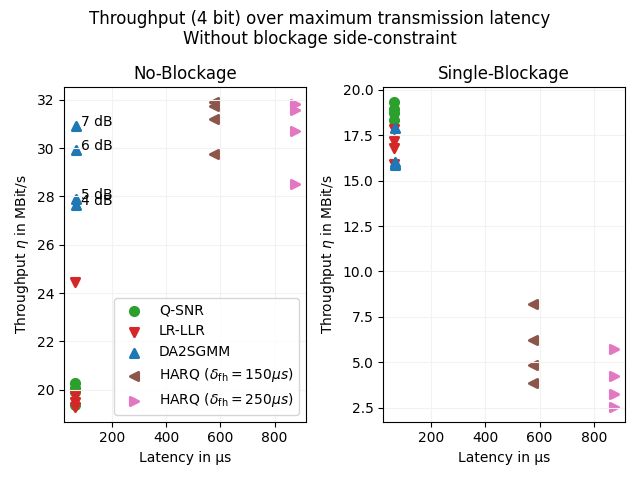}%
        \label{fig:harq_perf_1}}
          \fi
    \caption{Throughput (4~bits feedback) over the maximum transmission latency ($\delta_\mathrm{fb} = \delta_\mathrm{RV}$) with and without the blockage side constraint at the evaluated \acp{SNR} and corresponding target error rates, as provided in Tab.~\ref{tab:lls_assump}. (a) With blockage constraint. (b) Without blockage constraint.}
        \label{fig:harq_perf}
    
\end{figure}
In Fig.~\ref{fig:harq_perf}, we show the \ac{HARQ} performance of the prediction schemes with 4~bits feedback size in terms of the throughput with and without the blockage side constraint as defined in (\ref{eq:prHARQ_opt}). We assume $T_\mathrm{blk} := 4$, which implies that a positive blockage detection results in 4 additionally requested \acp{RV}. Furthermore, we set $\delta_\mathrm{fb} = \delta_\mathrm{RV}$. Due to the downlink control signaling design, i.e. PDCCH in 5G, it is possible to achieve even smaller feedback delays. However, as it is clear from (\ref{eq:lat_early}), any smaller $\delta_\mathrm{fb}$ would diminish the impact of the complexity differences of the schemes. For the processing latency of the schemes, we take the measurement results from the single-threaded implementation on an Intel Xeon CPU as a basis. In Fig.~\ref{fig:harq_perf_0}, we observe in the no-blockage scenarios that \ac{DIDA} with 23 - 25 Mbit/s throughput clearly outperforms all other prediction schemes, which achieve approximately 8 MBit/s at all \acp{SNR}. We note that the latency that is required to achieve the target error rates, does not differ significantly for the \ac{HARQ} prediction schemes. Compared to regular \ac{HARQ}, \ac{DIDA} reaches approximately 20~\% less throughput. Nevertheless, the higher throughput of regular \ac{HARQ} comes at the cost of a significantly larger maximum transmission latency compared to \ac{DIDA}. In the single-blockage scenario, we note that the additional latency due to retransmissions significantly degrades the throughput of regular \ac{HARQ}. The \ac{DIDA}, \ac{Q-SNR} and \ac{LR-LLR} achieve a similar and significantly higher throughput for all \acp{SNR} except the \unit[7]{dB} \ac{SNR} of \ac{HARQ} with $\delta_\mathrm{fh} = 150 \mu s$.
In Fig.~\ref{fig:harq_perf_1}, we show the throughput without the blockage side constraint. We observe that the performance improves for all prediction schemes in both scenarios. Especially, the \ac{Q-SNR} and \ac{LR-LLR} significantly benefit from removing this side constraint. This indicates that in the previous performance evaluations, these schemes are mainly limited by the stringent blockage detection side constraint defined in (\ref{eq:prHARQ_opt}).\\
Another critical issue for machine learning schemes is the robustness against unknown channel variations. In particular, any learned scheme has to reliably perform for a larger range of channel parameters, such as the \ac{SNR}. In order to test the robustness of the trained \ac{DIDA}, we exclude the \ac{SNR} of \unit[6]{dB} from training. We note that the throughput of \ac{DIDA} behaves as expected also at this \ac{SNR} point. Furthermore, we note that the achieved throughput at \unit[6]{dB} is even closer to the throughput of \unit[7]{dB} than \unit[5]{dB}, which hints that the scheme behaves as expected also in unknown channel variations.


\section{Summary and Conclusions}
In this work, we proposed novel machine-learning assisted \ac{HARQ} prediction schemes and evaluated them within the context of a \ac{C-RAN} scenario in the sub-THz regime consider also blockage using link-level simulations. In particular, we extended the \ac{LLR}- and subcode-based approaches proposed in \cite{journal_eharq_paper} enabling their usage in a \ac{C-RAN} setup by introducing quantization and a feedback combination module using a logistic regression and we proposed a novel end-to-end \ac{DIDA} architecture that exploits \ac{SNR} and subcode features. Using realistic link-level simulations, we showed that the proposed \ac{DIDA} clearly outperforms other prediction mechanisms in no-blockage as well as in single-blockage scenarios within the context of the \ac{HARQ} evaluation methodology. In particular, we present that the \ac{DIDA} \ac{HARQ} prediction achieves a more than 200~\% higher throughput compared to other \ac{HARQ} prediction schemes at \acp{SNR} ranging from \unit[4]{dB} to \unit[7]{dB} and target error rates from $1\cdot 10^{-4}$ to $3.5\cdot 10^{-6}$, if single-blockage is considered. Even without blockage, we show that the throughput of the \ac{DIDA} is approx. 29~\% higher compared to the \ac{LR-LLR} and even 45~\% higher compared to the \ac{Q-SNR}. Compared to regular \ac{HARQ}, our proposed \ac{DIDA} with a sufficient feedback size suffers only by a throughput reduction of approx. 20~\% while reducing the maximum transmission latency by a factor larger than 4. Furthermore, we show that 4 bits for the feedback transmission is sufficient and the schemes do not benefit from more bits. In future research, the impact of double-blockage as well as a setup with more than two \acp{RRH} may be studied. Furthermore, the impact of non-ideal channel estimation on the performance of the different schemes has to be evaluated in further studies.

\bibliographystyle{IEEEtran}
\bibliography{lib}

\appendices

\section{Dual-Input Denoising Autoencoder}\label{sec:dida_details}

The network configuration for the encoders at each \ac{RRH} is [FCL($d$,25), FCL(25,10), FCL(10,3)] and for the decoder [FCL(3,15), FCL(15,40), Lin(40,$d$)], where  FCL(x,y) $\equiv$ [Lin(x,y), BN, L-ReLU] and $d$ is the input dimension. Furthermore, Lin(x,y) denotes a linear transformation layer, BN a Batch Normalization-layer, L-ReLU a Leaky ReLU activation layer with a slope of $0.01$. 
Each \ac{RRH} further contains a classifier that operates on the compressed form of the subcode features. The network configurations of the \ac{RRH} classifiers each read as [FC(5,10), FC(10,15), FC(15, 15), FC(15, 10), Lin(10, 2), SM], where  FC(x,y) $\equiv$ [Lin(x,y), BN, ReLU] where ReLU is a ReLU activation layer and SM is a softmax activation layer. The classifiers each receive the compressed representation of $\overline{\boldsymbol{s}}^{(i)}$ and the received \ac{SNR} $\gamma^{(i)}$ as input. To prevent drifting off of the two "arms" of the network, the local encoders and classifiers are tied together. In particular this means, the weights and the biases of the linear layers are updated equally at both \acp{RRH}. Furthermore, we implement the quantization layer by a FakeQuantize layer with a MinMaxObserver in PyTorch using quantization-aware training \cite{quantization_pytorch}. Lastly, the network configuration of the \ac{UE} classifier is given by [FC(2,20), FC(20,10), FC(10,5), Lin(5,2), SM]. We train the \ac{DIDA} in an end-to-end fashion using a loss function $L$ that is composed by the $L_2$ norm:
\begin{equation}
    L_2 = \frac{1}{N} \sum_{i=1}^2 \sum_{k=1}^N||\boldsymbol{s}^{(i)}_k - \hat{\boldsymbol{s}}^{(i)}_k||_2^2\,,
\end{equation}
and the cross-entropy between the predicted output $\hat{d}$ and the actual decoding outcome $d$:
\begin{equation}
    L_\mathrm{ce} = \frac{1}{1 + \omega_\mathrm{ACK}} \left( d_k \log(\hat{d}_k) + \omega_\mathrm{ACK}(1 - d_k) \log(1 - \hat{d}_k)\right)\,,
\end{equation}
where $N$ is the number of samples in a batch\ and $\omega_\mathrm{ACK}$ is a weight class for \acp{ACK}. The two loss functions are combined as:
\begin{equation}
    L = L_2 + \lambda L_\mathrm{ce}\,,
\end{equation}
where $\lambda$ is a fixed weight factor. The fixed weight factor and the batch size are found to give the best results at $15.0$ and $15,000$, respectively, for all prediction points. For the first and the blockage prediction point, the \ac{ACK} weight class $\omega_\mathrm{ACK}$ achieves at $0.5$ the best performance. For the second prediction point, the \ac{ACK} weight class is chosen to be $0.1$. To train \ac{DIDA} under the given loss function, we use the Adam optimizer \cite{Kingma2014AdamAM} at a learning rate of 0.001 and weight decay of $10^{-5}$. We initialize the parameters of the whole network with the Kaiming normal initialization \cite{kaiming_init}. Due to the nature of the sample data, the ratio between ACKs and NACKs is heavily imbalanced. Hence, we undersample the majority class, i.e. ACKs, to create a balance between ACKs and NACKs.

\end{document}

%% file: acro.tex
\usepackage{acro}

\DeclareAcronym{URLLC}{
  short = URLLC ,
  long  = Ultra-Reliable Low Latency Communication ,
  class = abbrev
}

\DeclareAcronym{SINR}{
    short = {SINR} ,
    long  = Signal to Interference plus Noise Ratio ,
    class = abbrev
}

\DeclareAcronym{LR}{
    short = {LR} ,
    long  = Logistic Regression ,
    class = abbrev
}

\DeclareAcronym{LOS}{
    short = {LOS} ,
    long  = Line-Of-Sight ,
    class = abbrev
}

\DeclareAcronym{NLOS}{
    short = {NLOS} ,
    long  = Non-Line-Of-Sight ,
    class = abbrev
}

\DeclareAcronym{TDL-D}{
    short = {TDL-D} ,
    long  = Tapped Delay Line D ,
    class = abbrev
}

\DeclareAcronym{TDL}{
    short = {TDL} ,
    long  = Tapped Delay Line ,
    class = abbrev
}

\DeclareAcronym{CDL}{
    short = {CDL} ,
    long  = Clustered Delay Line ,
    class = abbrev
}

\DeclareAcronym{CDL-D}{
    short = {CDL-D} ,
    long  = Clustered Delay Line D ,
    class = abbrev
}

\DeclareAcronym{MDS}{
    short = {MDS} ,
    long  = Maximum Distance Separable ,
    class = abbrev
}

\DeclareAcronym{LR-LLR}{
    short = {LR-LLR} ,
    long  = Logistic Regression on LogLikelihood Ratios ,
    class = abbrev
}

\DeclareAcronym{AUC}{
    short = {AUC} ,
    long  = Area Under the false-positive-Curve ,
    class = abbrev
}

\DeclareAcronym{SLSQP}{
    short = {SLSQP} ,
    long  = Sequential Least Squares Programming ,
    class = abbrev
}

\DeclareAcronym{TH-SNR}{
    short = {LR-SNR} ,
    long  = Logistic Regression on Signal-to-Noise-Ratio ,
    class = abbrev
}

\DeclareAcronym{Q-SNR}{
    short = {Q-SNR} ,
    long  = Quantized Signal-to-Noise-Ratio ,
    class = abbrev
}

\DeclareAcronym{LR-SC}{
    short = {LR-SC} ,
    long  = Logistic Regression on Subcode features ,
    class = abbrev
}

\DeclareAcronym{TH-LLR}{
    short = {Q-LLR} ,
    long  = Quantized LogLikelihood Ratio ,
    class = abbrev
}

\DeclareAcronym{DIDA}{
    short = {DA2SGMM} ,
    long  = Dual Autoencoding 2-Stage Gaussian Mixture Model ,
    class = abbrev
}

\DeclareAcronym{IR}{
    short = {IR} ,
    long  = Incremental Redundancy ,
    class = abbrev
}

\DeclareAcronym{EEG}{
    short = {EEG} ,
    long  = Energy Efficiency Gain ,
    class = abbrev
}

\DeclareAcronym{DMRS}{
  short = DMRS ,
  long  = Demodulation Reference Signal,
  class = abbrev
}

\DeclareAcronym{WI}{
  short = WI ,
  long  = Work Item,
  class = abbrev
}

\DeclareAcronym{QAM}{
  short = QAM ,
  long  = Quadrature Amplitude Modulation ,
  class = abbrev
}

\DeclareAcronym{OFDM}{
  short = OFDM ,
  long  = Orthogonal Frequency Division Multiplexing ,
  class = abbrev
}

\DeclareAcronym{OFDMA}{
  short = OFDMA ,
  long  = Orthogonal Frequency Division Multiplexing Access ,
  class = abbrev
}

\DeclareAcronym{E2E}{
  short = E2E ,
  long  = End-to-End ,
  class = abbrev
}

\DeclareAcronym{DL}{
  short = DL ,
  long  = DownLink ,
  class = abbrev
}

\DeclareAcronym{PDCP}{
  short = PDCP ,
  long  = Packet Data Convergence Protocol ,
  class = abbrev
}

\DeclareAcronym{prHARQ}{
  short = prHARQ ,
  long  = proactive HARQ with prediction ,
  class = abbrev
}

\DeclareAcronym{eprHARQ}{
  short = eprHARQ ,
  long  = early proactive HARQ with prediction ,
  class = abbrev
}

\DeclareAcronym{paHARQ}{
  short = paHARQ ,
  long  = proactive HARQ ,
  class = abbrev
}

\DeclareAcronym{reHARQ}{
  short = reHARQ ,
  long  = reactive HARQ ,
  class = abbrev
}

\DeclareAcronym{GF}{
  short = GF ,
  long  = Grant-Free ,
  class = abbrev
}

\DeclareAcronym{ML}{
  short = ML ,
  long  = Machine Learning ,
  class = abbrev
}

\DeclareAcronym{RB}{
  short = RB ,
  long  = Resource Block ,
  class = abbrev
}

\DeclareAcronym{RAN}{
  short = RAN ,
  long  = Radio Access Network ,
  class = abbrev
}

\DeclareAcronym{CSI}{
  short = CSI ,
  long  = Channel State Information ,
  class = abbrev
}

\DeclareAcronym{CSIT}{
  short = CSIT ,
  long  = Channel State Information at the Transmitter ,
  class = abbrev
}

\DeclareAcronym{CSIR}{
  short = CSIR ,
  long  = Channel State Information at the Receiver ,
  class = abbrev
}

\DeclareAcronym{MCS}{
  short = MCS ,
  long  = Modulation and Coding Scheme ,
  class = abbrev
}

\DeclareAcronym{CFI}{
  short = CFI ,
  long  = Control Format Indicator ,
  class = abbrev
}

\DeclareAcronym{UL}{
  short = UL ,
  long  = UpLink ,
  class = abbrev
}

\DeclareAcronym{SR}{
  short = SR ,
  long  = Scheduling Request ,
  class = abbrev
}

\DeclareAcronym{BER}{
  short = BER ,
  long  = Bit Error Rate ,
  class = abbrev
}

\DeclareAcronym{BLER}{
  short = BLER ,
  long  = Block Error Rate ,
  class = abbrev
}

\DeclareAcronym{CRC}{
  short = CRC ,
  long  = Cyclic Redundancy Check ,
  class = abbrev
}

\DeclareAcronym{RRH}{
  short = RRH ,
  long  = Remote Radio Head ,
  class = abbrev
}

\DeclareAcronym{BBU}{
  short = BBU ,
  long  = BaseBand Unit ,
  class = abbrev
}

\DeclareAcronym{CB}{
  short = CB ,
  long  = Code Block ,
  class = abbrev
}

\DeclareAcronym{CBG}{
  short = CBG ,
  long  = Code Block Group ,
  class = abbrev
}

\DeclareAcronym{R-CBG}{
  short = R-CBG ,
  long  = Reduced Code Block Group ,
  class = abbrev
}

\DeclareAcronym{AR-CBG}{
  short = AR-CBG ,
  long  = Adaptive Reduced Code Block Group ,
  class = abbrev
}

\DeclareAcronym{TB}{
  short = TB ,
  long  = Transport Block ,
  class = abbrev
}

\DeclareAcronym{BG2}{
  short = BG2 ,
  long  = Base Graph 2 ,
  class = abbrev
}

\DeclareAcronym{BG1}{
  short = BG1 ,
  long  = Base Graph 1 ,
  class = abbrev
}

\DeclareAcronym{SNR}{
  short = SNR ,
  long  = Signal-to-Noise Ratio ,
  class = abbrev
}

\DeclareAcronym{CCE}{
  short = CCE ,
  long  = Control Channel Element ,
  class = abbrev
}

\DeclareAcronym{PDCCH}{
  short = PDCCH ,
  long  = Physical Downlink Control Channel ,
  class = abbrev
}

\DeclareAcronym{PUCCH}{
  short = PUCCH ,
  long  = Physical Uplink Control Channel ,
  class = abbrev
}

\DeclareAcronym{LTE}{
  short = LTE ,
  long  = Long Term Evolution ,
  class = abbrev
}

\DeclareAcronym{NGMN}{
  short = NGMN ,
  long  = Next Generation Mobile Networks ,
  class = abbrev
}

\DeclareAcronym{RNTI}{
  short = RNTI ,
  long  = Radio Network Temporary Identifier ,
  class = abbrev
}

\DeclareAcronym{FC}{
  short = FC ,
  long  = Fully Connected ,
  class = abbrev
}

\DeclareAcronym{3GPP}{
  short = 3GPP ,
  long  = 3rd Generation Partnership Project ,
  class = abbrev
}

\DeclareAcronym{HRLLC}{
  short = HRLLC ,
  long  = High-Reliable Low Latency Communication ,
  class = abbrev
}

\DeclareAcronym{SIMO}{
  short = SIMO ,
  long  = single-input multiple-output ,
  class = abbrev
}

\DeclareAcronym{MIMO}{
  short = MIMO ,
  long  = Multiple-Input Multiple-Output ,
  class = abbrev
}

\DeclareAcronym{TTI}{
  short = TTI ,
  long  = Transmission Time Interval ,
  class = abbrev
}

\DeclareAcronym{sTTI}{
  short = sTTI ,
  long  = short Transmission Time Interval ,
  class = abbrev
}

\DeclareAcronym{RTT}{
  short = RTT ,
  long  = Round Trip Time ,
  class = abbrev
}

\DeclareAcronym{LDPC}{
  short = LDPC ,
  long  = Low-Density Parity-Check ,
  class = abbrev
}

\DeclareAcronym{UE}{
  short = UE ,
  long  = User Equipment ,
  class = abbrev
}

\DeclareAcronym{TI}{
    short = TI ,
    long  = Tactile Internet ,
    class = abbrev
}

\DeclareAcronym{BS}{
  short = BS ,
  long  = Base Station ,
  class = abbrev
}

\DeclareAcronym{FPR}{
  short = FPR ,
  long  = False-Positive Rate ,
  class = abbrev
}

\DeclareAcronym{FNR}{
  short = FNR ,
  long  = False-Negative Rate ,
  class = abbrev
}

\DeclareAcronym{DCI}{
  short = DCI ,
  long  = Downlink Control Information ,
  class = abbrev
}

\DeclareAcronym{HARQ}{
  short = HARQ ,
  long  = Hybrid Automatic Repeat reQuest ,
  class = abbrev
}

\DeclareAcronym{IIOT}{
  short = IIOT ,
  long  = Industrial Internet of Things ,
  class = abbrev
}

\DeclareAcronym{RV}{
  short = RV ,
  long  = Redundancy Version ,
  class = abbrev
}

\DeclareAcronym{ACK}{
  short = ACK ,
  long  = ACKnowledgment ,
  class = abbrev
}

\DeclareAcronym{NACK}{
  short = NACK ,
  long  = Non-ACKnowledgment ,
  class = abbrev
}

\DeclareAcronym{CG}{
  short = CG ,
  long  = Configured Grant ,
  class = abbrev
}

\DeclareAcronym{M-TRP}{
  short = M-TRP ,
  long  = Multi-TRP ,
  class = abbrev
}

\DeclareAcronym{E-HARQ}{
  short = E-HARQ ,
  long  = Early HARQ ,
  class = abbrev
}

\DeclareAcronym{P-HARQ}{
  short = P-HARQ ,
  long  = Predictive incremental-redundancy rateless HARQ ,
  class = abbrev
}

\DeclareAcronym{R-HARQ}{
  short = R-HARQ ,
  long  = Rateless incremental-redundancy HARQ ,
  class = abbrev
}

\DeclareAcronym{C-RAN}{
  short = C-RAN ,
  long  = Cloud Radio Access Network ,
  class = abbrev
}

\DeclareAcronym{C-HARQ}{
  short = C-HARQ ,
  long  = Conventional incremental-redundancy HARQ ,
  class = abbrev
}

\DeclareAcronym{mMTC}{
  short = mMTC ,
  long  = massive Machine Type Communications ,
  class = abbrev
}

\DeclareAcronym{5G}{
  short = 5G ,
  long  = Fifth Generation ,
  class = abbrev
}

\DeclareAcronym{6G}{
  short = 6G ,
  long  = Sixth Generation ,
  class = abbrev
}

\DeclareAcronym{SPS}{
  short = SPS ,
  long  = Semi-Persistent Scheduling ,
  class = abbrev
}

\DeclareAcronym{PI}{
  short = PI ,
  long  = Pre-emption Indication ,
  class = abbrev
}

\DeclareAcronym{V2X}{
  short = V2X ,
  long  = Vehicle-To-Everything ,
  class = abbrev
}

\DeclareAcronym{VR}{
  short = VR ,
  long  = Virtual Reality ,
  class = abbrev
}

\DeclareAcronym{NR}{
  short = NR ,
  long  = New Radio ,
  class = abbrev
}

\DeclareAcronym{eMBB}{
  short = eMBB ,
  long  = enhanced Mobile BroadBand ,
  class = abbrev
}

\DeclareAcronym{LLR}{
  short = LLR ,
  long  = Log-Likelihood Ratio ,
  class = abbrev
}

\DeclareAcronym{VNR}{
  short = VNR ,
  long  = Variable Node Reliability ,
  class = abbrev
}

\DeclareAcronym{BSC}{
  short = BSC ,
  long  = Binary Symmetric Channel ,
  class = abbrev
}

\DeclareAcronym{angelsperarea}{
  short = \ensuremath{a} ,
  long  = The number of angels per unit area ,
  sort  = a ,
  class = nomencl
}
\DeclareAcronym{numofangels}{
  short = \ensuremath{N} ,
  long  = The number of angels per needle point ,
  sort  = N ,
  class = nomencl
}
\DeclareAcronym{areaofneedle}{
  short = \ensuremath{A} ,
  long  = The area of the needle point ,
  sort  = A ,
  class = nomencl
}